# A fractal geometry enhanced topology optimization design for high-performance liquid cooling plates

Zixu Han, Kairan Yang, Peng Zhang*

Institute of Refrigeration and Cryogenics, Shanghai Jiao Tong University, Shanghai 200240, China

**Abstract**

The density-based bi-objective topology optimization (TO) has been widely adopted in liquid cooling plate design, where the design domain is treated as porous media with porosity as the design variable. However, conventional TO method struggles to directly optimize the convective heat transfer due to its incapabilities of explicitly depicting the heat transfer area in objective function, which limits the optimization of thermal performance. In this study, a fractal geometry topology optimization (FGTO) method is proposed, which incorporates fractal dimension as an additional design freedom into the density-based TO framework. Different from the conventional TO methods, the FGTO explicitly describes the heat transfer area, and achieves a direct optimization of convective heat transfer through the objective function. Compared to the conventional TO, the FGTO achieves a more complex structural topology in the optimized liquid cooling plate with a 46% improvement in heat transfer area. The fractal dimension is manipulated by varying the input parameter *s*, and increasing *s* can improve thermal performance of the FGTO results at the cost of larger pressure drop. Superior thermal-hydraulic performance can be achieved by varying *s*, with the average and maximum temperatures of the FGTO results reduced by 15.6 K and 16.9 K, respectively, compared with those of the conventional TO results. The integration of fractal geometry into the TO intensifies the difference in objective function sensitivity between solid and liquid phases, which is conducive to facilitating solid-liquid separation and contributes to escape from local optimal solutions.

**Keywords:** Topology optimization; Fractal geometry; Thermal management; Liquid cooling plate

**Nomenclature**

---

* Corresponding author: zhangp@sjtu.edu.cn

| | | | | |
|---|---|---|---|---|
| $A$ | specific area, m$^{-1}$ | | $n$ | pores number density, m$^{-2}$ |
| $A_p$ | projection area, m$^2$ | | $N$ | box number |
| $A_s$ | heat transfer area, m$^2$ | | $Nu$ | Nusselt number |
| $c_1$ | perimeter parameter | | $p$ | pressure, Pa |
| $c_2$ | area parameter | | $\Delta P$ | pressure drop, Pa |
| $c_3$ | constant parameter | | $q$ | penalty factor, heat flux, W/m$^2$ |
| $C$ | perimeter, m | | $Q$ | heat source, W/m$^3$ |
| $C_p$ | specific heat capacity, J/(kg·K) | | $r$ | filter radius, m |
| $D_f$ | fractal dimension of porous media | | $R$ | equality constrain |
| $D_T$ | fractal dimension of liquid cooling plates | | $Re$ | Reynolds number |
| $Da$ | Darcy number | | $s$ | maximum-minimum pore size ratio |
| $h$ | heat transfer coefficient, W/(m$^2$·K) | | $T$ | temperature, K |
| $h^*$ | local heat transfer coefficient, W/(m$^2$·K) | | $\hat{u}$ | velocity, m/s |
| $k$ | thermal conductivity, W/(m·K) | | $V_{f1}$ | minimum liquid fraction |
| $K$ | solid-liquid conductivity ratio | | $V_{f2}$ | maximum liquid fraction |
| $l$ | characteristic length, m | | $w$ | weight parameter |

*Greek letters*

| | | | | |
|---|---|---|---|---|
| $\alpha$ | Brinkman friction coefficient | | $\gamma$ | design variable |
| $\delta$ | thickness, m | | $\lambda$ | pore size, m |
| $\varepsilon$ | Sensitivity of objective function | | $\sigma$ | box size |
| $\varphi$ | local objective function | | $\sigma^*$ | generalized heat transfer coefficient, W/(m$^2$·K) |
| $\Psi$ | objective function | | | |

*Acronyms*

| | | | | |
|---|---|---|---|---|
| FGTO | fractal geometry topology optimization | | RCP | rectangular cooling plate |
| PEC | performance evaluation criterion | | TO | topology optimization |

*Subscripts*

| | | | | |
|---|---|---|---|---|
| $a$ | adjoint variable | $out$ | outlet |
| $\beta$ | projection point | $p$ | projection |
| $f$ | fluid | $s$ | solid |
| $in$ | inlet | $t$ | thermal |
| $m$ | average | $tc$ | thermal objective by conventional method |
| $max$ | maximum | $0$ | initial value |
| $min$ | minimum | | |

# 1. Introduction

The trend toward miniaturization and integration of electronic devices inevitably leads to higher heat density, which significantly deteriorate the performance and lifespan of electronic devices, while compromising the reliability [1]. Therefore, effective and efficient thermal management gradually becomes one of the most critical challenges in various scenarios. Forced convection air cooling systems are traditionally used in the thermal management, but their heat dissipation capabilities are limited and cannot meet the increasing demand for high heat flux cooling [2]. Consequently, liquid cooling, including direct and indirect contact cooling, has been widely adopted. Direct contact cooling, also known as immersion cooling, poses high sealing requirements for the cooling devices and the coolant is costly, which limits its application [3]. Currently, indirect contact cooling through liquid cooling plates is widely used due to high heat dissipation capability, low cost, high flexibility and reliability [4]. Therefore, effective optimization of the thermal and hydraulic performances of liquid cooling plates is crucial for enhancing the performance and reliability of electronic devices.

Thermal and hydraulic demands, geometric structures, and operating conditions vary significantly for different electronic components. For example, lithium-ion batteries require temperature control within a specific range of 15-35°C, with heightened demands for temperature uniformity [5]. Simultaneously, the liquid cooling plate must minimize flow resistance to ensure energy efficiency of the battery system, demonstrating demand for high hydraulic performance. For power electronic components with high power density, like the insulated-gate bipolar transistor (IGBT), temperatures must be maintained below 200 °C, presenting stringent thermal requirements and minimal hydraulic demands [6]. For CPUs/GPUs, temperatures must be kept below 70 °C while reducing pumping power consumption, necessitating moderate thermal and hydraulic requirements [7]. Consequently, for different scenarios, liquid cooling plate design must be tailored to various thermal and hydraulic demands and operating conditions [8], placing additional requirements on the flexibility of design and optimization method. However, traditional design method of liquid cooling plates significantly relies on adjusting single geometric parameter, e.g., channel width or fin size [9], and requires extensive numerical endeavors, leading to the failure in achieving efficient liquid cooling plate design. Therefore, it is necessary to develop a more flexible optimization method, which can efficiently manipulate structures and performances of the liquid cooling plates.

Topology optimization (TO) is a non-intuitive design method that permits a high degree of design freedom in structures and is capable of producing an innovative optimal design with minimal reliance on the intuition and experience of the designer, which can enhance the universality and efficiency of design optimization [10]. In recent years, TO has been increasingly attractive in thermo–fluidic investigations, and demonstrated superior performance over traditional optimization methods in liquid cooling plates design [11]. Most of the current TOs are using the density-based method proposed by Bendsøe and Kikuchi [12], which treats design domain as a porous medium and optimizes the porosity distribution in the design domain. Since then, the density-based TO method has been continuously refined to meet the requirements for optimizing liquid cooling plates [13-16].

Zeng and Lee [17] proposed a two-layer TO model with one layer of solid substrate and the other of fluid and achieved a 50% reduction in pumping power compared to the straight channels. Yang et al. [18] employed a three-layer TO model to optimize the heat transfer between hot and cold fluids mediated by a solid wall, and the resulting TO design achieved a 10.8% increase in heat dissipation rate compared to airfoil structures, while simultaneously reducing pumping power by 17%. Compared with single-layer TO, the pseudo-3D multi-layer TO achieved a higher fidelity but required more optimization parameters and higher computational costs. For the complex thermo-fluidic TO problems, the heavy computational costs might cause TO to become more prone to converge to local optimal solutions [19]. Consequently, the multi-layer TO models might underperform the simplified single-layer one in some cases [20], and most TO studies for liquid cooling plates are performed using the single-layer model. Liu et al. [21] adopted a bi-objective TO targeting heat dissipation rate and pumping power for liquid cooling plate design, achieving a 33% improvement in heat transfer coefficient and 70% reduction in pressure drop compared to the serpentine cold plate. Wu et al. [22] introduced temperature variance as an additional constraint to the bi-objective TO to improve thermal uniformity of liquid cooling plates, achieving a maximum temperature reduction of 1.8 K compared to the RCP. Tang et al. [23] employed average temperature and temperature variance as the thermal objectives for TO, reducing maximum temperature by 17.0 K and 13.6 K, respectively, compared to the RCP. Xia et al. [24] conducted bi-objective TO under various inlet/outlet configurations and found that the flared inlet/outlet structure achieved the best overall performance, with pumping power reduced by 20%-50% compared to the conventional straight structure while maintaining comparable thermal performance. Zou et al. [25] studied the effects of the thermal weight on the bi-objective TO results and found that the optimization became inefficient and converged poorly when the thermal weight exceeded 0.9. Sun et al. [26] conducted bi-objective TO for the liquid cooling plates under multi-inlets/outlets configurations. They found that increasing thermal weight from 0.6 to 0.8 reduced average temperature by only 2.1 K while the pressure drop was increased by about 50%. Consequently, the optimization of thermal performance was compromised and the computational costs increased rapidly at high thermal weights. Wang et al. [27] conducted the bi-objective TO using various thermal objectives, including the average solid temperature, heat dissipation rate, and the outlet fluid enthalpy, and they found that the selection of thermal objective demonstrated significant impacts on TO and the structural topology of the optimized liquid cooling plates could be manipulated through varying thermal objectives.

Since both thermal and hydraulic performances should be considered in liquid cooling plate design, most of the TO studies targeted at bi-objective optimization. Thermal and hydraulic objectives are correlated through the simple weighted-sum method and performance of the TO results is manipulated by varying the thermal weight. However, under extreme thermal weight, i.e., when the weight approaches to 0 or 1, computational stability is compromised, and intermediate regions with poor convergence and even impractical structures like blockage might emerge, which hinder practical applications and degrade performance of the TO results [24, 28]. Moreover, for the complex non-convex optimization problems like thermo-fluidic coupling multi-objective TO, it is not accessible to obtain the complete Pareto frontier, i.e., the global optimal solution [29, 30]. Especially when the weight approaches unity or zero, where one objective dominates the counterpart, the optimization becomes inefficient, requiring extreme sacrifices in hydraulic performance to achieve a slight improvement in thermal performance, and even fails to further enhance the thermal performance of optimized liquid cooling plates [25-27]. Consequently, thermal performance optimization by the conventional TO is limited by inefficient performance manipulation approach of simply increasing thermal weight, and it struggles to adapt to scenarios with extremely high thermal demand, such as the power electronic components.

Wang et al. [31] proposed an epsilon-constraint TO method that replaces the weighted-sum approach with multiple constraint equations to achieve more efficient bi-objective TO. Their approach enhanced the computational stability and efficiency of the TO and significantly reduced the intermediate regions with poor convergence. However, compared to the weighted-sum approach, their approach could only improve the convergence of the TO but had little effect on the macroscopic structural topology of the TO results. Some TO studies incorporated hydraulic objectives such as pressure drop or pumping power through constraint equations rather than objective functions in the TO, enabling optimization for a single thermal objective [32, 33]. However, the effects of additional constraints and objective functions on the TO are essentially similar, while unreasonable constraints would even severely impair convergence, resulting in an impractical structural topology [22]. Therefore, the issue of limited thermal performance optimization by the conventional bi-objective TO remains unresolved.

The optimization of thermal performance by the conventional bi-objective TO not only suffers from intrinsic limitations of the weighted-sum approach, but it is also constrained by the insufficient model integrity of the optimized porous media. The density-based TO treats the design domain as a porous medium and porosity as the design variable. It optimizes the porosity distribution based on the sensitivity of objective function, and then the optimized porous media is facilitated to converge to pure solid or liquid phases through approaches like interpolation, density filtering and projection. The objective function affects the TO process through the sensitivity, and different objective functions would lead to various evolutions of the porosity distributions that correspond to the improved objectives, ultimately resulting in different structural topologies. Since heat dissipation of liquid cooling plate depends primarily on convective heat transfer, the optimal thermal objective should directly target maximizing the thermal convection, requiring the lateral heat transfer area to be explicitly depicted. However, since only the porosity is considered in the conventional TO, it struggles to reflect sufficient topological features in optimization and fails to explicitly depict the heat transfer area of the porous media. Consequently, the conventional TO could only optimize convective heat transfer indirectly, e.g., minimizing average temperature [34] or maximizing the heat transfer by heat conduction [21], which hardly reflect the main heat transfer mechanism of liquid cooling plate, and constrain the augmentation of heat transfer area and optimization of thermal performance.

The integration of interface tracking techniques into the TOs is one of the effective approaches to addressing the aforementioned issues, e.g., spline parameterizations or level-set based TO methods. The former describes structural topologies using spline curves, treating control points as design variables to control the solid-liquid interface directly [35]. However, such approach is limited to adjusting fin shapes, restricting design freedom, and necessitates mesh reconstruction after each iteration, thereby increasing computational complexity [36]. In contrast, the level-set based TO describes structural topologies by updating a signed distance function. This approach retains advantages on performance improvement of density-based TO while it can achieve sharper and clearer solid-liquid interfaces and avoid the zigzag boundary issue in density-based TO [37]. However, this method requires solving an additional governing equation for the signed distance function, leading to a significant increase in computational time when coupled with the nonlinear thermo-fluidic problems. Furthermore, additional constraints on minimum geometric dimensions are required in the level-set TO to ensure a practical optimized structure, and the mesh dependency issues also limit its application [37, 38]. In contrast, density-based TO methods can more readily constrain minimum geometric scales and resolve mesh dependency issues through filtering and projection techniques, while the update of design variable does not require additional governing equations. Therefore, refining the porous medium modeling and objective function design within the density-based TO framework, remains the most effective and feasible approach for further enhancing thermal performance of the optimized liquid cooling plates.

Fractal geometry theory can mathematically describe the topological characteristics of complex structures, like tree leaves or respiratory systems through the fractal dimension, and has been applied to depict topological features of porous media [39, 40] or branched flow channels [41]. Recently, the fractal geometry theory has been increasingly employed to the optimization of heat sinks or fins. Wang et al. [42] obtained different microchannel structures by varying the fractal dimension, then further optimized these structures through the TO. Dong et al. [43] combined bionic optimization with the TO. They firstly evaluated and selected heat sinks with leaf-like topological features using fractal dimension analysis, then further optimized the selected bionic structures by the TO. Zhao et al. [44] and Song et al. [45] employed the conventional TO to obtain bionic fin structures with excellent heat transfer performance, then utilized fractal dimension to quantitatively evaluate their topological characteristics. These studies adopted the fractal dimension as a criterion for evaluating topological features of the optimized structures, rather than incorporating it directly into the TO process as a design freedom.

To the best of the authors' knowledge, there are no existing studies on incorporating the fractal dimension as an independent design freedom into the thermo-fluidic TO problems. In this study, a fractal geometry topology optimization (FGTO) method is proposed, which integrates the fractal geometry theory into the density-based TO framework through an objective function design and manipulate the optimization results by varying the input parameter $s$. The primary difference between the FGTO and conventional TO lies in its capability of explicitly describing heat transfer area by fractal geometry theory, and it directly optimizes convective heat transfer through its objective function. Compared to the conventional TO, the optimized liquid cooling plate by the FGTO achieves more complex solid and flow channel structures, and a significantly augmented heat transfer area, leading to the reduced average and maximum temperatures, and the improved performance evaluation criterion (PEC) of the optimized liquid cooling plates, which is further discussed in section 3. Overall, the FGTO refines the model integrity of optimized porous media in the conventional density-based TO framework, and therefore enhances the upper limits of thermal performance optimization for liquid cooling plates.

## 2. FGTO model

### 2.1 Model geometry and boundary conditions

**Fig. 1(a)** illustrates several typical application scenarios for liquid cooling plates, including power electronic components with high thermal loads, CPUs/GPUs with moderate thermal and hydraulic demands, and power batteries with high energy efficiency and hydraulic demands. The liquid cooling plate is modeled independently to achieve a versatile design and performance manipulation across scenarios with varying thermal and hydraulic demands, as shown in **Fig. 1(b)**. The solid phase of liquid cooling plates is copper, and the coolant is ethylene glycol, with thermo-physical properties listed in **Table 1**. The inlet coolant flows at 0.2 m/s with a temperature of 308.15 K, and the outlet boundary is set as a pressure outlet. In practical applications, heat sources might be non-uniform or time-varying. However, this study focuses on the discussion of a new TO methodology, prioritizing the performance comparison of various liquid cooling plates. Additionally, the liquid cooling plates are operated under single-phase laminar flow condition for brevity. Consequently, the heat source is simplified to a constant and uniform heat flux of 10 W/cm² applied to the heated plate, as shown in **Fig. 1(b)**.

High Reynolds numbers or large heat fluxes would compromise computational stability and degrade performance of the TO results [27, 46]. Therefore, most TOs are conducted under more moderate conditions with relatively low *Re* [47]. Consequently, a simplified 2D model is adopted for the FGTO, as shown in **Fig. 1(c)**, with the inlet flow velocity of 0.01 m/s, inlet temperature of 308.15 K and heat flux of 1 W/cm². The real operating boundary conditions shown in **Fig. 1(b)** are applied to the 3D numerical calculations for evaluating performance of the optimized liquid cooling plates by the FGTO method. The grid number of 2,015,217 is used for the 3D numerical calculations, and further details about the model settings and mesh-independent check for the 3D numerical calculations can be found in **Section 1** in the **supplementary materials**.

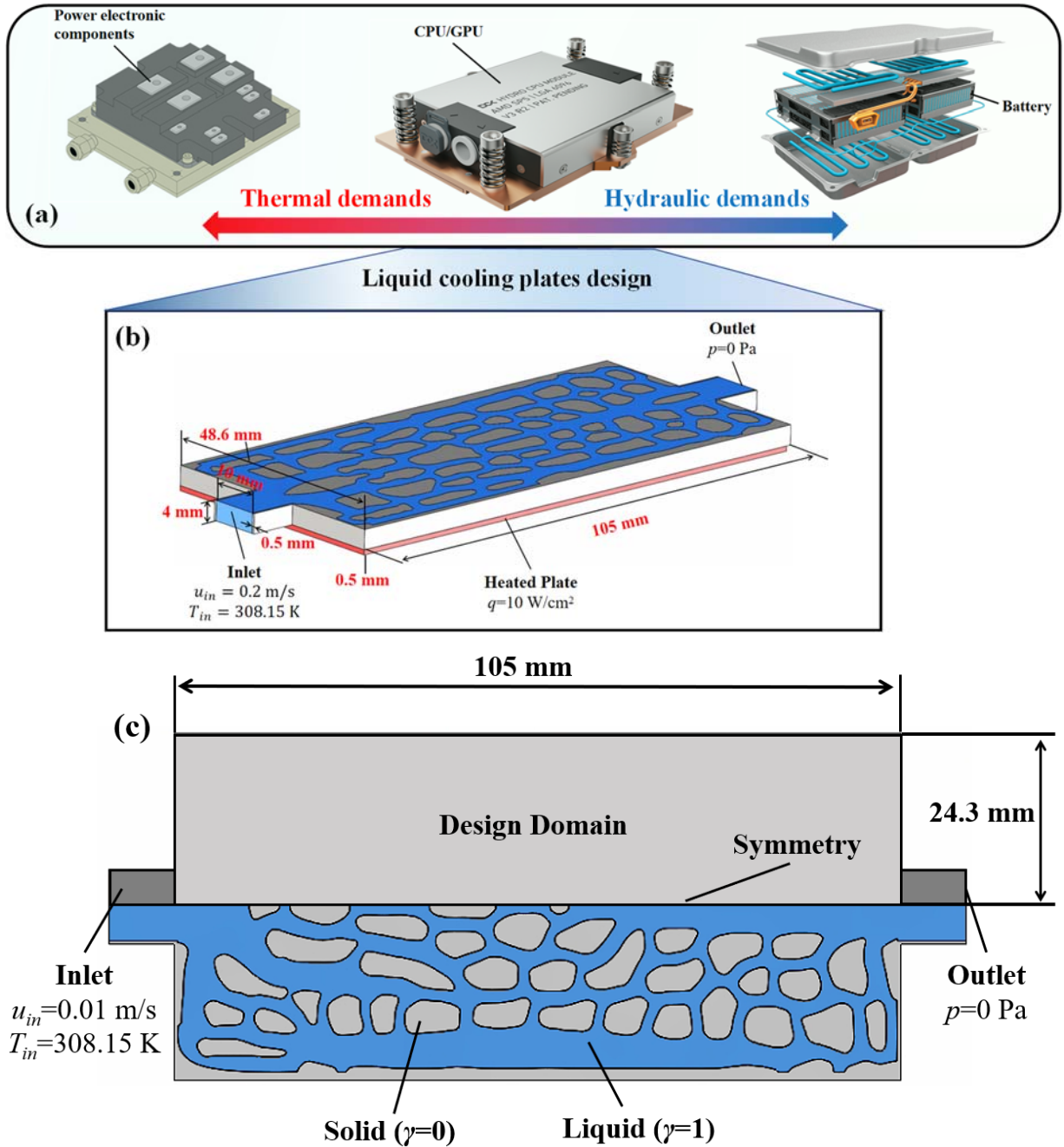

**Fig. 1.** The FGTO model for liquid cooling plate design and performance manipulation. (a) application scenarios for liquid cooling plates, including power electronic components, CPUs/GPUs, and batteries [48-50], (b) schematic of the 3D numerical calculation model for performance evaluation, (c) schematic of the 2D FGTO model.

**Table 1** Thermo-physical properties of the solid and fluid.

| Material | $\rho$ [kg/m³] | $C_p$ [J/(kg·K)] | $k$ [W/(m·K)] | $\mu$ [Pa·s] |
|---|---|---|---|---|
| Ethylene glycol | 1049 | 3454 | 0.4 | 0.00128 |
| Copper | 8900 | 385 | 385 | - |

## 2.2 Fractal geometry theory and objective function design

The density-based TO method treats design domain as a porous medium and porosity $\gamma$ as the design variable [47]. **Fig. 2 (a)** illustrates schematic of the density-based TO and different approaches to modeling the porous media by the conventional TO and FGTO methods. The porosity distribution within the design domain is optimized through sensitivity analysis. Then the solid ($\gamma=0$) and liquid ($\gamma=1$) phases are separated from the optimized porous media ($0<\gamma<1$) through density filtering and projection, while the detailed topological features are eliminated for practical applications, as shown in **Fig. 2(a)**. In such a TO process, the optimized liquid cooling plate can be considered as an approximate solution of the optimal porous medium structure, and the convergence from the porous medium to liquid cooling plate can be regarded as a process of pore coalescence, as shown in **Fig. 2(a)**. Therefore, performance optimization by the density-base TO depends on the evolution process of porosity distribution.

The porosity, i.e., the design variable, is updated based on the sensitivity analysis of objective function, so the evolution of porosity distribution and the ultimate TO results are significantly influenced by the design of thermal objectives [27]. Since heat dissipation of liquid cooling plates primarily relies on the thermal convection, the optimal thermal objective for liquid cooling plate design should directly target maximizing the convective heat transfer, requiring an explicit description of the lateral heat transfer area. However, since only the porosity is considered in the conventional TO, it fails to explicitly depict heat transfer area of the optimized porous media and therefore does not reflect the characteristics of convective heat transfer in thermal objective function. Consequently, the conventional TO can only optimize thermal convection indirectly, e.g., minimizing average temperature [34], maximizing heat transfer by heat conduction [51] or outlet fluid enthalpy [27], constraining the improvement of convective heat transfer performance. To improve performance of the TO results, the FGTO refines the porous medium model by incorporating other topological features of porous media beyond porosity $\gamma$ through an additional design freedom of fractal dimension $D_f$, as shown in **Fig. 2(a)**.

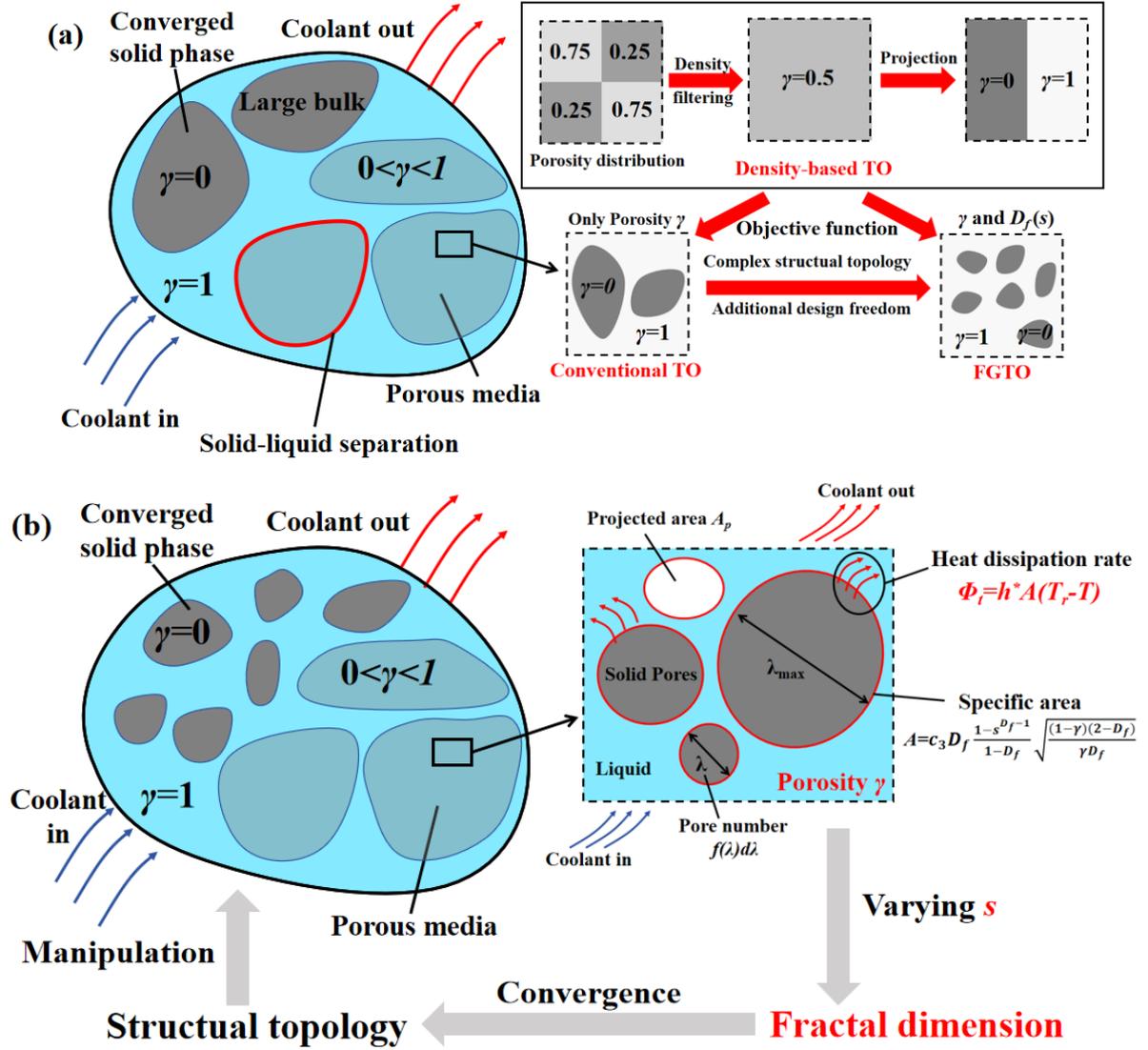

**Fig. 2.** Schematic of the conventional TO and FGTO methods. (a) conventional TO method, (b) the FGTO method incorporating porous medium model by fractal geometry theory.

**Fig. 2(b)** shows the schematic of the FGTO method including modeling the porous medium based on fractal geometry theory, which can explicitly depict specific area of the optimized porous media. The effect of fractal dimension $D_f$ is fully reflected in the objective function through the depicted heat transfer area. Therefore, the FGTO can also be regarded as an optimization framework targeted at achieving a fractal dimension, whereby the design domain gradually evolves from a homogeneous porous medium toward a structural topology with a predetermined fractal dimension. By varying the input parameter $s$ according to Eq. (1), one can modify the predefined targeted fractal dimension, thereby affecting the variation scaling of heat transfer area with $\gamma$ within the design domain, resulting in different structural topologies in the optimization results. Increasing $s$ can enhance the expected fractal dimension in the objective function, thereby resulting in liquid cooling plates with augmented heat transfer area and more complex structures, leading to an improved thermal performance, which is further discussed in section 3.

The relationship between local fractal dimension and porosity of microscopic porous medium is given by [39]:

$$D_f = D_E + \frac{\ln(1-\gamma)}{\ln(\frac{\lambda_{max}}{\lambda_{min}})} = D_E + \frac{\ln(1-\gamma)}{\ln(s)} \tag{1}$$

where $D_f$ is the fractal dimension of porous media with a porosity of $\gamma$, which represents the complexity of the local microscopic structures. $s=\frac{\lambda_{max}}{\lambda_{min}}$ is the ratio of maximum pore size to minimum one. $D_E$ is the Euclidean dimension with 2 and 3 for 2D and 3D dimensions, respectively. Since $\gamma$ continuously varies in the optimization process, it is inconvenient to directly modify the $D_f$. Therefore, the fractal dimension is indirectly adjusted by varying the parameter $s$. It should be noted that the parameter $s$, is different from the design variable $\gamma$ that continuously varies in the design domain, and it serves as an input parameter that has a clear qualitative correlation to the structure and performance of the optimized liquid cooling plate, which is further discussed in section 3. The self-similarity assumption of the fractal geometry theory requires $s$ to take a large value, e.g., $10^2$-$10^4$ is recommended for the porous media [39].

As the topological features at each geometric scale exhibit the same characteristics due to self-similarity, an area of unit size is chosen as the study region for convenience. The total number of pores in the studied region is expressed as [52]:

$$n = s^{D_f} \tag{2}$$

The probability density function of the pore size distribution can be formulated as [53]:

$$f(\lambda) = D_f \left(\frac{\lambda_{max}}{s}\right)^{D_f} \lambda^{-(D_f+1)} \tag{3}$$

The perimeter of the porous structure in the studied region can be expressed as:

$$C = \int_{\lambda_{min}}^{\lambda_{max}} nf(\lambda) c_1 \lambda d\lambda = c_1 D_f \lambda_{max} \frac{1-s^{D_f-1}}{1-D_f} \tag{4}$$

where $c_1$ is a non-dimensional shape factor whose product of the geometric scale of the aperture is equal to the perimeter of a single aperture, e.g., $c_1= \pi$ for a circle shape and $c_1=4$ for a square shape. For a 2D case, the total projected area of the porous structure along the thickness direction can be expressed as:

$$A_p = \int_{\lambda_{min}}^{\lambda_{max}} nf(\lambda)c_2\lambda^2 d\lambda = c_2\lambda_{max}^2 \frac{D_f}{2-D_f}(1-s^{D_f-2}) = c_2\gamma\lambda_{max}^2 \frac{D_f}{2-D_f} \tag{5}$$

where $c_2$ is a non-dimensional shape factor whose product with the squared geometric scale is equal to the projected area of a single pore structure, e.g., $c_2 = \pi/4$ for a circle shape and $c_2=1$ for a square shape. The projected area of the studied region with unit area can also be formulated as:

$$A_p = 1[m] \times 1[m] \times (1-\gamma) = 1-\gamma \tag{6}$$

The maximum pore size can be yielded as follows using Eqs. (5) and (6):

$$\lambda_{max} = \sqrt{\frac{1-\gamma}{\frac{D_f}{2-D_f}\gamma c_2}} \tag{7}$$

Therefore, the total lateral heat transfer area of the liquid cooling plate can be estimated by $A_s = C\delta$, with $\delta = 4$ mm being thickness of the liquid cooling plate. The expression for the heat transfer area per volume $A$, i.e., the specific surface area, can be obtained by coupling Eqs. (1)-(4) and (7):

$$A = \frac{A_s}{1[m] \times 1[m] \times \delta} = c_3 D_f \frac{1-s^{D_f-1}}{1-D_f} \sqrt{\frac{1-\gamma}{\frac{D_f}{2-D_f}\gamma}} \tag{8}$$

where $c_3 = \frac{c_1}{\sqrt{c_2}}$ is a constant which represents compactness of apertures with the assumed shape, and it is independent of the geometry of design domain, design variable and other state variables in the FGTO. The thermal objective function characterizing the total heat dissipation by thermal convection in the design domain can be formulated as:

$$\Psi_t = \int_\Omega \phi_t(\gamma) \, d\Omega = \int_\Omega h^* A(T_r - T) \, d\Omega = \int_\Omega c_3 h^* D_f \frac{1-s^{D_f-1}}{1-D_f} \sqrt{\frac{1-\gamma}{\frac{D_f}{2-D_f}\gamma}} (T_r - T) \, d\Omega \tag{9}$$

where $\phi_t(\gamma)$ is local thermal objective function, $\Omega$ is the design domain region, $T_r$ is the reference temperature of the heat source, and $h^*$ is the local heat transfer coefficient. Since the heat is transferred from solid to fluid, $T_r$ represents the solid temperature while $T$ represents the local fluid temperature. Consequently, the term $T_r$-$T$ in Eq. (9) reflects the local temperature difference in convective heat transfer. Due to the continuous evolution of the optimized structural topology through iterations, it is extremely difficult to accurately calculate and update the local heat transfer coefficient during the TO process. Therefore, many previous TO studies also simplified such "generalized heat transfer coefficient" to a constant, while their objective functions did not explicitly depict the heat transfer area [21, 51]. Since this study focuses on incorporating the fractal geometry into the density-based TO framework to explicitly depict the heat transfer area, the spatially varying $h^*$ is simplified as a constant in this study.

As the constants $c_3$ and $h^*$ are not relevant to the design variable and other physical quantities, these constants would be eliminated in the dimensionless transformation of the objective function, i.e., the selection of these constants would not influence the evolution of the design variable by affecting the sensitivity of the objective function. Consequently, the thermal objective function in Eq. (9) can be simplified as $\Psi_t = \int_\Omega D_f \frac{1-s^{D_f-1}}{1-D_f} \sqrt{\frac{1-\gamma}{\frac{D_f}{2-D_f}\gamma}} (T_r - T) \, d\Omega$ in the FGTO. As shown in **Fig. 2(b),** by varying $s$, i.e., $D_f$ in Eq. (1), the objective function Eq. (9) can be modulated, which in turn affects the evolution of the porosity distribution, i.e., the design variable, during the FGTO process. Therefore, the structural topology of the FGTO results can be manipulated through varying $s$.

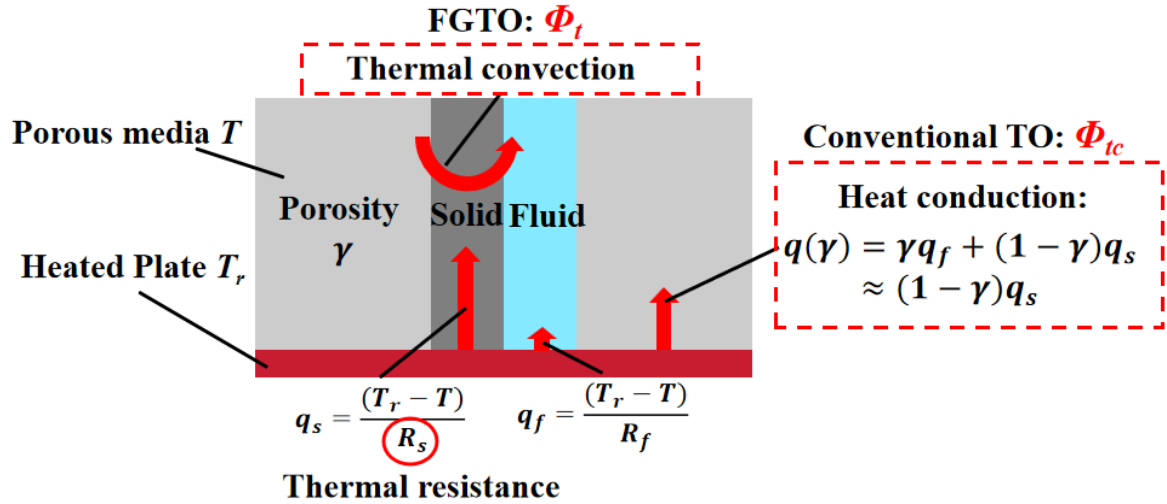

**Fig. 3.** Comparison of the thermal objectives by the FGTO and conventional TO. The FGTO targets maximizing heat dissipation by convective heat transfer through lateral heat transfer area in porous media, while the conventional TO targets maximizing heat dissipation by heat conduction.

**Fig. 3** illustrates the distinction in the thermal objectives by the FGTO and conventional TO. The fractal geometry theory enables the FGTO to explicitly depicts the lateral area for convective heat transfer of the optimized porous media and to directly target maximizing heat dissipation by thermal convection, as shown in Eq. (9). Conversely, only porosity is considered in the conventional TO, it targets maximizing heat dissipation by heat conduction from the heated plate to the porous media, which can be formulated as $q(\gamma)=\gamma q_f+(1-\gamma)\, q_s$, with $q_f$ and $q_s$ the heat conduction in pure fluid and solid, respectively. By assuming the thermal resistance of solids is much smaller than that of fluids, it can be simplified to the most commonly used thermal objective by conventional TO as [21, 51]:

$$\Psi_{tc} = \int_{\Omega} \phi_{tc}(\gamma)\, d\Omega = \int_{\Omega} \sigma^*(1-\gamma)(T_r - T)\, d\Omega \tag{10}$$

where $\phi_{tc}(\gamma)$ is local thermal objective function by the conventional TO method, and the constant $\sigma^*$ is a generalized heat transfer coefficient, which actually represents thermal conductance of solid. Such an essential heat transfer distinction in thermal objectives of the FGTO and conventional TO leads to the distinctive thermal performances of the optimized liquid cooling plates.

### 2.3 The box-counting method

In order to quantitatively evaluate the complexity of the optimized structures, the box-counting method [54] is used to estimate the overall fractal dimension of the optimized liquid cooling plates. By identifying the solid and liquid areas in the image of the FGTO results, the full image is divided into boxes of a certain size $\sigma$. The total number of boxes in the solid area $N(\sigma)$ is counted, and the above process is repeated with different box sizes. Finally, the fractal dimension of the FGTO results $D_T$, which represents structural complexity of the optimized liquid cooling plates, can be formulated as [43]:

$$D_T = -\lim_{\sigma \to 0} \frac{\lg(N(\sigma))}{\lg(\sigma)} \tag{11}$$

### 2.4 Governing equations

By assuming the steady incompressible laminar flow of the coolant in liquid cooling plate, the flow field is depicted by the continuity equation and Navier-Stokes equation:

$$R_p = -\nabla \cdot (\rho_f \hat{u}) \tag{12}$$

$$R_u = \nabla \cdot (\rho_f \hat{u}\hat{u}) - \nabla \cdot \mu(\nabla \hat{u} + \nabla \hat{u}^T) + \nabla p + \alpha(\gamma)\hat{u} \tag{13}$$

where $R$ is the equality constraint, and the Brinkman friction coefficient $\alpha(\gamma) = \frac{\mu}{Da l^2} \frac{q(1-\gamma)}{q+\gamma}$ is calculated by the Darcy interpolation function [55].

Considering the design domain as a porous medium, the constraint representation of the energy equation is given as [21]:

$$R_T = \gamma \rho_f C_{pf}(\hat{u} \cdot \nabla)T - [(1-\gamma)k_s + \gamma k_f]\nabla \cdot (\nabla T) - (1-\gamma)Q \tag{14}$$

where $Q$ is the heat source and the subscripts s and f represent the solid and fluid, respectively. For the bi-objective problem, the thermal objective function is derived from Eq. (9), and the flow dissipation in the entire domains $\Omega$ is chosen as hydraulic objective function to represent flow resistance [13]:

$$\Psi_f = \int_\Omega \phi_f(\gamma)\, d\Omega = \int_\Omega (\mu \nabla \hat{u} \cdot \nabla \hat{u} + \alpha(\gamma)\hat{u} \cdot \hat{u})\, d\Omega \tag{15}$$

where $\phi_f(\gamma)$ is the local hydraulic objective function. The objective functions of Eqs. (9) and (15) are normalized and weighted to obtain the final objective function as follows:

$$\Psi = w_t \Psi^*_t - w_f \Psi^*_f \tag{16}$$

where $w_t$ and $w_f$ are the weights of the dimensionless thermal objective $\Psi^*_t = \frac{\Psi_t}{\Psi_{t0}}$ and dimensionless hydraulic objective $\Psi^*_f = \frac{\Psi_f}{\Psi_{f0}}$, respectively, with $w_t + w_f = 1$. $\Psi_{t0}$, $\Psi_{f0}$ are the initial values of $\Psi_t$ and $\Psi_f$, respectively. To achieve better overall performance, the objective function of Eq. (16) should be maximized.

The Helmholtz-type partial differential equation density filter is applied to avoid checkerboard problems and to obtain mesh-independent results [56]:

$$\gamma_f = r^2 \nabla^2 \gamma_f + \gamma \tag{17}$$

where $\gamma_f$ is the filtered design variable; and $r$ is the filter radius that determines the minimum length scale of the optimized structures, which is set as 2.5 × mesh size in this study. The filter inevitably introduces an intermediate-density area between the solid and fluid, which can be reduced by a hyperbolic tangent projection [57]:

$$\gamma_p = \frac{\tanh(\beta(\gamma_f - \gamma_\beta)) + \tanh(\beta \gamma_\beta)}{\tanh(\beta(1 - \gamma_\beta)) + \tanh(\beta \gamma_\beta)} \tag{18}$$

where $\beta$ is the slope, $\gamma_\beta$ is the projection point, and $\gamma_p$ is the projected design variable. In summary, the bi-objective optimization of liquid cooling plates by the FGTO can be formulated as:

$$\begin{aligned} Maximize: \quad & w_t \frac{\Psi_t(\gamma, s)}{\Psi_{t0}} - w_f \frac{\Psi_f}{\Psi_{f0}} \\ Subject\ to \quad & \begin{cases} R_p, R_u, R_T \\ V_{f1} \leq \dfrac{\int \gamma\, d\Omega}{\int d\Omega} \leq V_{f2} \\ w_t + w_f = 1\ (0 \leq w_t \leq 1) \\ 10^2 \leq s \leq 10^4 \end{cases} \end{aligned} \tag{19}$$

where the $V_{f1}$ and $V_{f2}$ are the upper and lower limits of the liquid fraction, respectively, which are set to accelerate convergence and ensure a practical optimized structural topology. The details about the above optimization parameters are listed in **Table 2**.

**Table 2** Values of the important parameters.

| Variable | Symbol | Values |
|---|---|---|
| Penalty parameter | $q$ | 0.01 |
| Darcy number | $Da$ | $10^{-4}$ |
| Filter radius | $r$ | 2.5 × mesh size |
| Projection slope | $\beta$ | 8 |
| Projection point | $\gamma_\beta$ | 0.5 |
| Minimum liquid fraction | $V_{f1}$ | 0.5 |
| Maximum liquid fraction | $V_{f2}$ | 0.7 |

## 2.5 Sensitivity analysis

To solve the above topology optimization problem, analyze the mechanism of the FGTO method and determine how various parameters and boundary conditions affect optimization, the sensitivity of the objective function with respect to the design variable is derived and computed based on the continuous adjoint method [55, 58, 59]. The Lagrange function is constructed as:

$$L = \Psi + \int_\Omega \hat{\lambda} \cdot \hat{R}(\hat{w},\gamma) d\Omega = \Psi + \langle \hat{\lambda}, \hat{R}(\hat{w},\gamma) \rangle \tag{20}$$

where $\hat{\lambda}=(p_a,\hat{u}_a,T_a)^\mathrm{T}$ is the Lagrange multiplier or adjoint variable, $\hat{w}=(p,\hat{u},T)^\mathrm{T}$ is the state variable vector and $\hat{R}(\hat{w},\gamma)=(R_p,R_u,R_T)^\mathrm{T}$ is the equality constraint vector. The sensitivity of the objective function with respect to the design variable is equivalent to the total derivative of the Lagrange function:

$$\begin{aligned}\frac{dL}{d\gamma} &= \frac{\partial \Psi}{\partial \gamma} + \frac{\partial \Psi}{\partial \hat{w}}\frac{d\hat{w}}{d\gamma} + \frac{\partial \langle \hat{\lambda}, \hat{R}(\hat{w},\gamma) \rangle}{\partial \gamma} + \frac{\partial \langle \hat{\lambda}, \hat{R}(\hat{w},\gamma) \rangle}{\partial \hat{w}}\frac{d\hat{w}}{d\gamma} \\ &= \frac{\partial \Psi}{\partial \gamma} + \frac{\partial \langle \hat{\lambda}, \hat{R}(\hat{w},\gamma) \rangle}{\partial \gamma} + \left( \frac{\partial \Psi}{\partial \hat{w}} + \frac{\partial \langle \hat{\lambda}, \hat{R}(\hat{w},\gamma) \rangle}{\partial \hat{w}} \right)\frac{d\hat{w}}{d\gamma}\end{aligned} \tag{21}$$

According to the KKT (Karush-Kuhn-Tucker) conditions for the PDE [60], the adjoint equation can be constructed as:

$$\frac{\partial \Psi}{\partial \hat{w}} + \frac{\partial \langle \hat{\lambda}, \hat{R}(\hat{w},\gamma) \rangle}{\partial \hat{w}} = 0 \tag{22}$$

Substituting Eq. (22) into Eq. (21), the Lagrange function can be simplified as:

$$\frac{dL}{d\gamma} = \frac{\partial \Psi}{\partial \gamma} + \frac{\partial \langle \hat{\lambda}, \hat{R}(\hat{w},\gamma) \rangle}{\partial \gamma} \tag{23}$$

The final expression for the sensitivity of the objective function is formulated as:

$$\frac{dL}{d\gamma} = w_t \frac{\int_\Omega \frac{\partial \phi_t(\gamma)}{\partial \gamma} d\Omega}{\psi_{t0}} - w_f \frac{\int_\Omega \frac{\partial \phi_f(\gamma)}{\partial \gamma} \hat{u} \cdot \hat{u} d\Omega}{\psi_{f0}} + \quad (24)$$

$$\int_\Omega (\frac{\partial \alpha(\gamma)}{\partial \gamma} \hat{u}_a \cdot \hat{u} + \rho C_{pf} T_a \hat{u} \cdot \nabla T + k_f (1 - \frac{k_s}{k_f}) \nabla T \cdot \nabla T_a + T_a Q) d\Omega$$

Eq. (24) can reflect the effects of thermal and hydraulic objective functions, objective weights, thermo-physical properties, solid-liquid thermal conductivity ratio and heat flux on the optimization. For simplicity, the dimensionless $\frac{\partial \phi_t(\gamma)}{\partial \gamma}$ in the first term of Eq. (24) is denoted as $\varepsilon_t$. The details of the specific derivation of Eq. (24) and the derivation of the adjoint variable governing equations can be found in **Section 2** in the **supplementary materials**.

Through sensitivity analysis, it is found that the mechanism by which the FGTO method promotes escaping from the local optimal solutions lies in increasing the differentiation in sensitivities between the solid and liquid phases. Increasing $s$ intensifies the sensitivity differentiation and makes the large solid bulks easier to break into smaller ones, thereby augmenting structural complexity of the optimized liquid cooling plate. According to the fractal geometry theory [39] and the influence of $s$ on the FGTO results, $s$ is recommended to range from $10^2$ to $10^4$, with more details discussed in Section 3.1.

## 2.6 Numerical methods

The topology optimization problem mentioned above are solved by COMSOL Multiphysics (6.0). First, a uniform initial design variable field of $\gamma$=0.5 is given. Second, the governing equations are solved based on the Finite Element Method (FEM). Third, the sensitivity analysis is conducted based on the methods in Section 2.5. Finally, the design variable is updated by the optimization solver based on the results of the sensitivity analysis. The Global Convergence Method of Moving Asymptotes optimization solver (GCMMA) [61] is adopted, and the simulation is terminated when the maximum residual of the design variable is less than $10^{-6}$. The COUPLED algorithm is adopted for solving the continuity and momentum equations in 3D numerical calculations. The calculation is terminated when the residuals of mass and velocity are less than $10^{-6}$, while the residual of energy is less than $10^{-8}$.

## 2.7 Mesh-independent analysis for the FGTO

As a larger mesh gird number is required to ensure computational accuracy with more complex structural topology of the optimized liquid cooling plate, the FGTO at $s$=8000 and $w_t$=0.7 with the highest structural complexity is selected as a representative condition for mesh-independent analysis. Structured square grids of identical dimensions are employed throughout the design domain, which ensures that the self-similarity across different geometric scales would not be compromised by the mesh structure. The mesh-independent check results of the 2D FGTO are presented in **Table 3**. It is found that the deviation in the objective function becomes unremarkable when the mesh grid number is greater than 37,928. Considering both computational cost and accuracy, a mesh grid number of 37,928 is ultimately used in the FGTO.

Table 3 The results of 2D mesh-independent check

| Grid number | $\Psi$ | Deviation |
| --- | --- | --- |
| 3352 | 0.6451 | 8.69% |
| 10251 | 0.7012 | 2.32% |
| 18798 | 0.7175 | 0.52% |
| 37928 | 0.7212 | 0.04% |
| 64401 | 0.7215 | - |

## 2.8 Verification of the numerical calculation

The 3D numerical calculations is verified by the experimental and numerical simulation results in [62]. The physical model, geometry structure and boundary conditions are consistent with those in [62]. The flow channel and fin widths are 2 mm, the *Re* ranges from 675 to 1013 and heat flux is around 9.35 W/cm$^2$, and more details can be found in Section 1.4 in **supplementary materials**. **Fig. 4** shows the comparison of the experimental and numerical results. It is demonstrated that maximum deviations between the experimental and numerical results for the average temperature of the heating surface and pressure drop are 0.7 K and 11.3%, respectively. Consequently, the 3D numerical calculation approach is considered reliable.

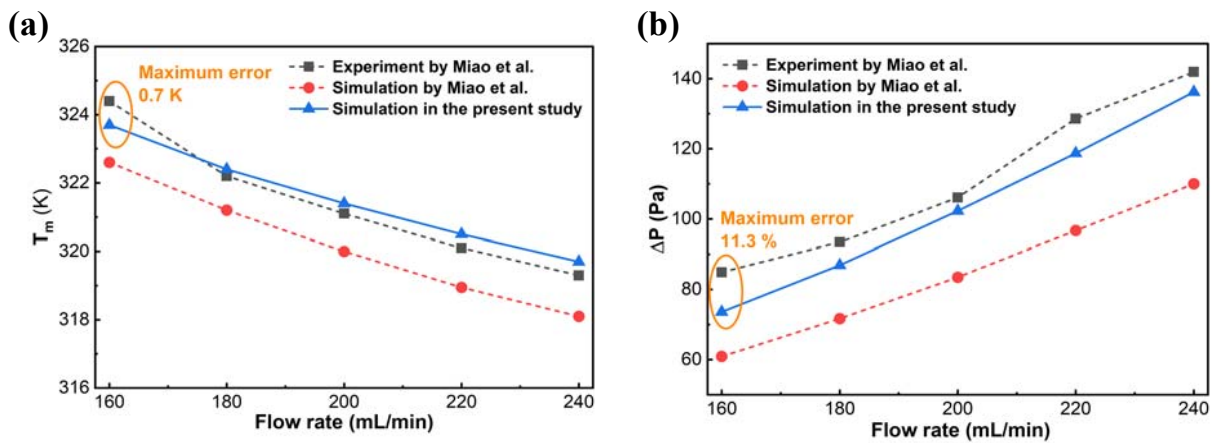

**Fig. 4.** Comparison of the experimental results in [62] and numerical results. (a) variation of average

temperature of the heating surface with flow rate, (b) variation of pressure drop with flow rate

## 3. Results and discussion

### 3.1 Effects of *s* on the FGTO results

**Fig. 5** shows the FGTO results at various *s* at $w_f$=0.7. By varying *s*, the fractal dimension within the design domain can be modified according to Eq. (1), influencing the evolution of design variables and consequently yielding different structural topologies of the optimized liquid cooling plates. It can be observed from **Fig. 5(a)** that increasing *s* makes bulk solid blocks more easily split into smaller ones, augmenting heat transfer area and tortuosity of the flow channels. As shown in **Table 4**, the heat transfer area is increased by 59% by increasing *s* from 1000 to 8000. Simultaneously, the fractal dimension $D_T$, i.e., the structural complexity of the optimized liquid cooling plate, also increases from 1.32 to 1.60, leading to a stronger flow perturbation and better heat transfer performance.

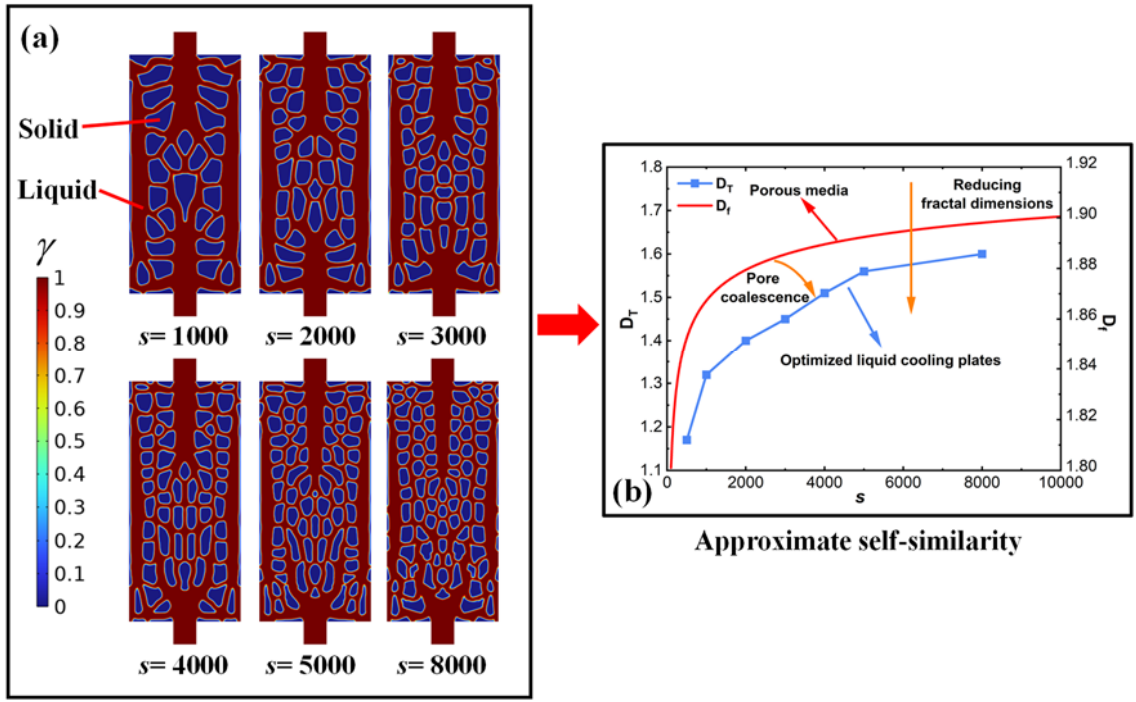

**Fig. 5.** The FGTO results at various *s* at $w_f$=0.7. (a) the design variable field of the FGTO result at *s*=1000, 2000, 3000, 4000, 5000, 8000, (b) variations of the fractal dimension of porous media $D_f$ and the overall fractal dimension of the optimized liquid cooling plates $D_T$ with *s*.

**Table 4** Overall fractal dimension and heat transfer area of different liquid cooling plates.

| Structure | RCP | *s*=1000 | *s*=3000 | *s*=5000 | *s*=8000 |
|---|---|---|---|---|---|
| $D_T$ | 1.08 | 1.32 | 1.45 | 1.56 | 1.60 |
| $A_s$ (mm$^2$) | 10123 | 4641.7 | 5681.8 | 6857.8 | 7380.8 |

Varying the input parameter $s$ can influence the predefined fractal dimension in the objective function, thereby causing the optimized porous medium to gradually converge toward structural topologies with the desired fractal dimension. However, due to the inherent complexity of the thermo-fluidic multi-objective TO problem, coupled with the fact that the density-based TO framework requires techniques such as filtering or projection to ensure a practical optimized structure, the fractal dimension of the optimized liquid cooling plate $D_T$ would be lower than the expected target $D_f$, as shown in **Fig. 5(b)**. Nevertheless, the fractal dimensions of $D_T$ and $D_f$ still exhibit the similar variation trends, where both gradually increase with rising $s$ before their growth rates level off. Consequently, the FGTO displays an approximate self-similarity, which demonstrates the effectiveness of varying the fractal dimension, i.e., the parameter $s$, to guide the evolution of the design variables to achieve liquid cooling plates with various structures and performances.

Since the FGTO enables explicit depiction and direct optimization of lateral heat transfer area by the thermal objective in Eq. (9), it provides a variation scaling of the heat transfer area with design variables across the design domain, which helps to more comprehensively reflect the mechanisms of convective heat transfer. Consequently, it is more conducive to solid-liquid separation compared to the conventional TO, leading to a larger heat transfer area in the optimized liquid cooling plate. The influence of thermal objective on the evolution of porosity distribution, i.e., on the TO process, is mathematically reflected through the sensitivity of objective function. It can be understood from Eq. (24) that the thermal objective function shows an impact on the sensitivity mainly through $\varepsilon_t$ in the first term on the right side of the equation. **Fig. 6** demonstrates the variation of $\varepsilon_t$ with the design variable at various $s$. It is demonstrated that the FGTO method accentuates the sensitivity differentiation between solid and liquid, thus facilitates the separation of solid and liquid phases from the optimized porous media. Therefore, the FGTO enables the generation of more complex structural topology. In contrast, sensitivity of the thermal objective function by the conventional TO in Eq. (10) remains constant with respect to the design variable, as shown in **Fig. 6**. Consequently, it fails to achieve sufficient topological features from the optimized porous media (0<$\gamma$<1).

Increasing $s$ intensifies the differentiation of sensitivity between solid and liquid phases, as shown in **Fig. 6**. For the FGTO at $s$=1000, 3000, and 8000, the sensitivity differences between solid and liquid phases are increased by 211%, 234%, and 251%, respectively, compared to the conventional thermal objective. Consequently, the solid and liquid phases are more readily separated and the complexity of structural topologies of the optimized liquid cooling plates is enhanced as *s* increases, as shown in **Fig. 5**.

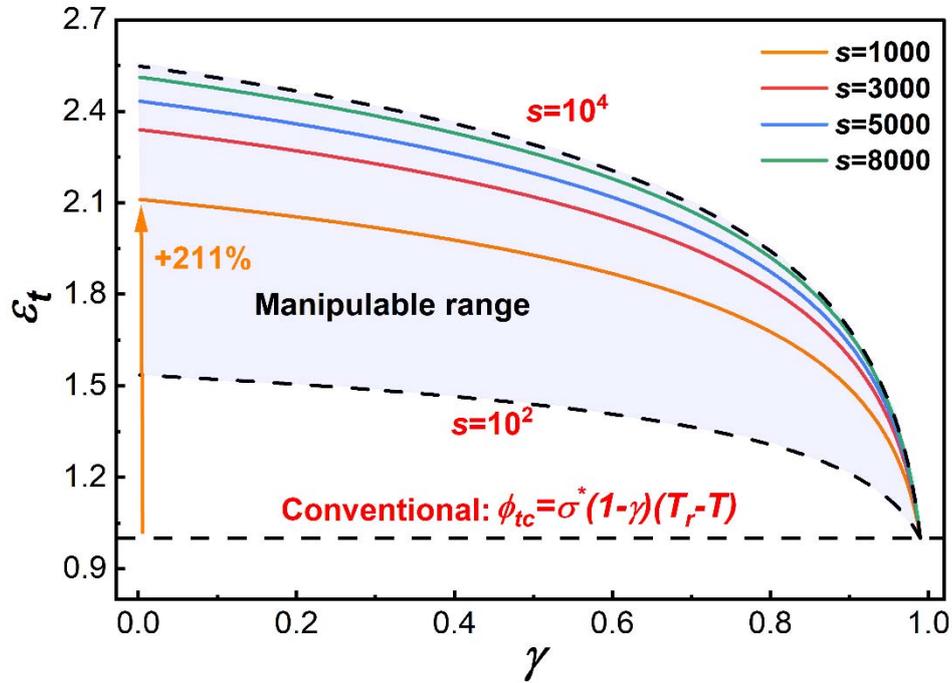

**Fig. 6.** Variation of the dimensionless sensitivity of thermal objective function $\varepsilon_t$ with $\gamma$ at different $s$

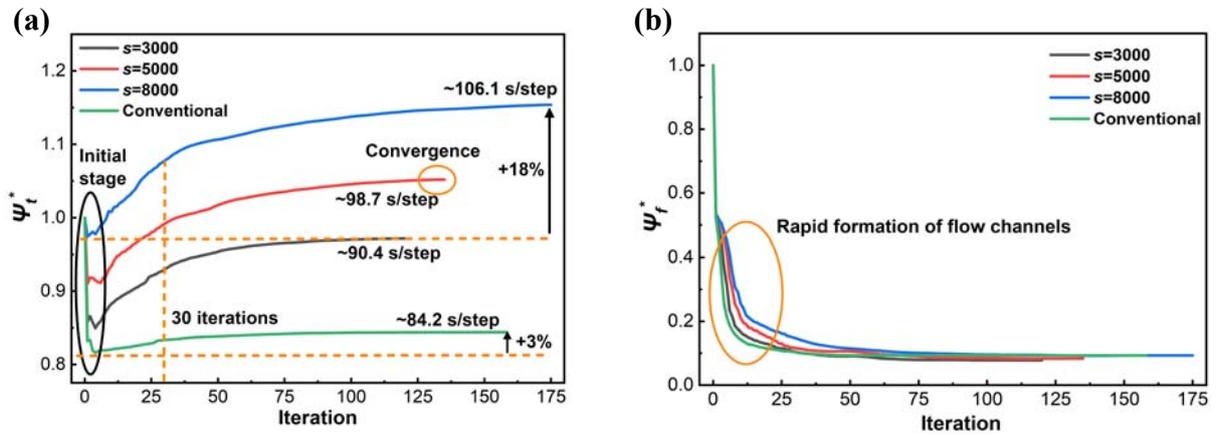

**Fig. 7.** Variation of the dimensionless objective functions with iteration steps by the FGTO at $w_t$=0.7 and $s$=3000, 5000, 8000, and the conventional TO at $w_t$=0.7. (a) dimensionless thermal objective function, (b) dimensionless hydraulic objective function

**Fig. 7(a)** and **Fig. 7(b)** show the variation of the dimensionless thermal and hydraulic objectives with iteration steps by the FGTO at $w_t$=0.7 and $s$=3000, 5000, 8000, and the conventional TO at $w_t$=0.7, respectively. To synchronously illustrate the formation of structural topology, **Fig. 8(a)** and **Fig. 8(b)** show the iteration and convergence processes of the conventional TO at $w_t$=0.7, and the FGTO at $s$=5000 and $w_t$=0.7, respectively. It can be observed from **Fig. 7** that the hydraulic objective is improved significantly while the thermal objective decreases rapidly in the initial several iterations for both the conventional TO and the FGTO, which corresponds to the process that the initial uniform porous medium rapidly converges to wide flow channels to reduce flow resistance, while relatively large solid bulks are generated simultaneously, as shown in **Fig. 8**. During the initial stage, the solution places excessive emphasis on improving hydraulic objective while despises the thermal objective, making it highly susceptible to falling into local optimal solutions with large solid bulks. Since the solid-liquid phase separation occurs more readily in the FGTO due to larger sensitivity difference between solid and liquid phases, narrower flow channels and smaller solid bulks are generated during the initial stage compared to the conventional TO, as shown in **Fig. 8(b)**. Therefore, the thermal objective of the FGTO exhibits less decay during the initial stage, as shown in **Fig. 7(a)**, indicating that the FGTO is more adept at avoiding being trapped at local optimal solutions. The FGTO at larger $s$ is more prone to achieve more complex structural topology, leading to a smaller reduction in the thermal objective, while slightly compromising in the hydraulic objective during the initial stage. As shown in **Fig. 7(a)**, the thermal objectives of the FGTO at $s$=3000, 5000, and 8000 decrease by 15.1%, 8.5%, and 2.3%, respectively, while that of the conventional TO decreases by 18.4%, suggesting that increasing $s$ contributes to avoid the local optimal solutions.

After the initial stage, both the thermal and hydraulic objectives are progressively optimized, while the improvement gradually levels off as iterations proceed until convergence is achieved. The conventional TO generates relatively simple structural topology with large solid bulks and its optimization of thermal objective saturates after only about 30 iteration steps. The initially formed large bulk solids struggle to partition into smaller pieces, while many regions with poor convergence are observed in the conventional TO, as circled in **Fig. 8(a)**. Consequently, the conventional TO requires much more iterative steps to eliminate these poorly converged regions with only a slight improvement of 3% in thermal objectives, as shown in **Fig. 7(a)**. Conversely, the initially formed solid is continuously fragmented into smaller pieces in the FGTO, i.e., it continuously escapes from local optimal solutions as iterations proceed, as shown in **Fig. 8(b)**. The design variable in the FGTO exhibits high sensitivity within the region with large solid bulks, as shown in **Fig. 6**, which makes local design variables more prone to change and escape from the local optimal solutions, leading to the formation of liquid spots within the solid bulks, as circled in **Fig. 8(b)**. Since the sensitivity gradually decreases from solid ($\gamma$=0) to liquid ($\gamma$=1) phases, as shown in **Fig. 6**, these newly formed liquid spots tend to remain stable. Therefore, the initial liquid spots gradually expand and converge to flow channels, breaking the large solid bulks into smaller ones. Such a process is repeated as iterations proceed during which thermal objectives of the FGTO are continuously refined. Thus, the FGTO ultimately achieves a fragmented solid structural topology in the optimized liquid cooling plate, which contributes to an augmented heat transfer area and enhanced flow perturbation. Consequently, the FGTO achieves a more significant improvement in thermal performance compared to the conventional TO, e.g., the thermal objective for the FGTO at $s$=8000 is improved by 18% compared to its minimal value, while only by 3% for the conventional TO, as shown in **Fig. 7(a)**.

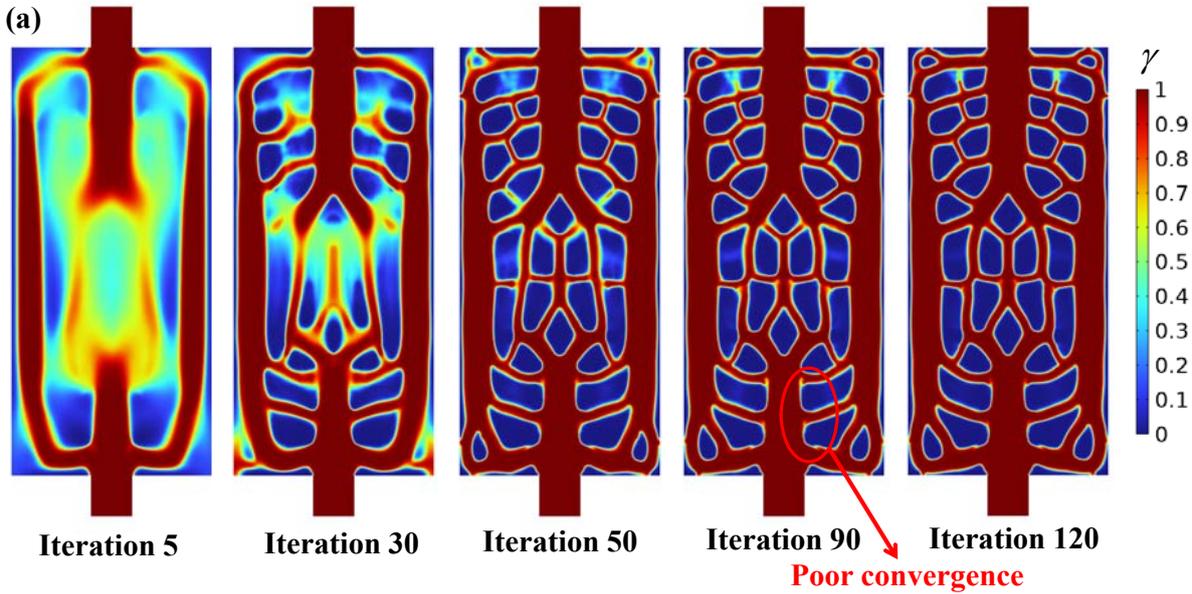

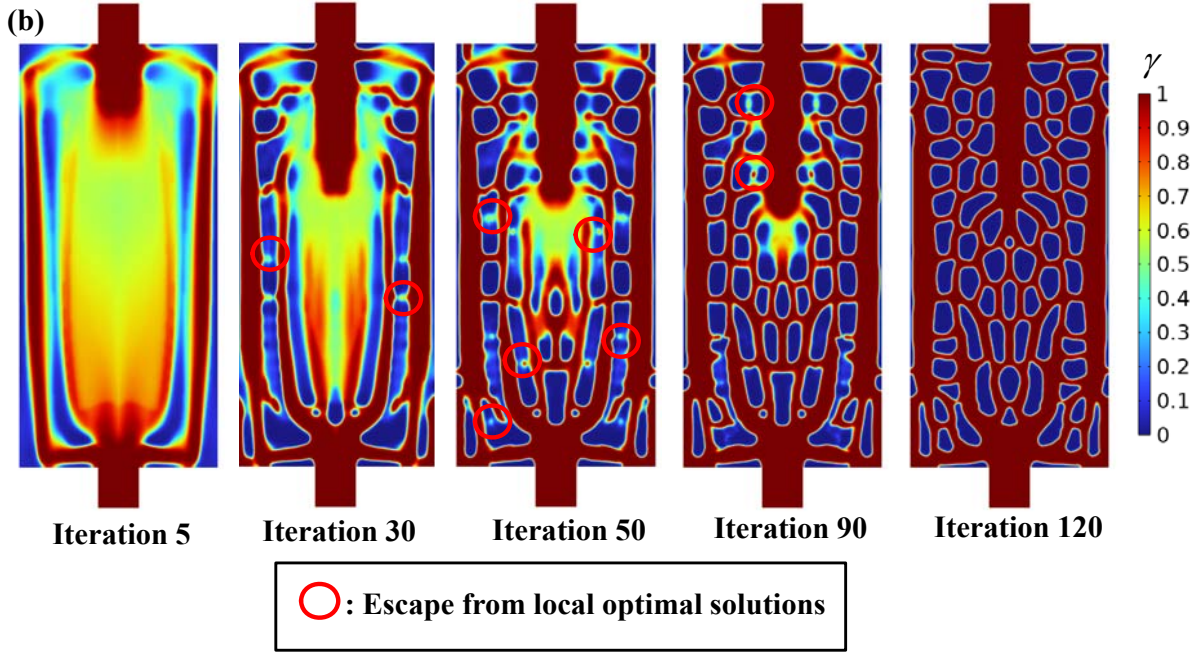

**Fig. 8.** Iteration and convergence processes of the FGTO and conventional TO. (a) the conventional TO at $w_t$=0.7, (b) the FGTO at $s$=5000 and $w_t$=0.7

**Table 5** Iteration steps and iteration time of the FGTO at $s$=3000, 5000, 8000, and the conventional TO at $w_t$=0.7.

|  | Conventional | $s$=3000 | $s$=5000 | $s$=8000 |
| --- | --- | --- | --- | --- |
| Iteration time per step (s) | 84.2 | 90.4 | 98.7 | 106.1 |
| Total steps | 159 | 120 | 135 | 176 |
| Total time (s) | 13,388 | 10,848 | 13,325 | 18,674 |

Overall, the FGTO method promotes separation between solid and liquid phases from the optimized porous media, leading to a higher structural complexity and more pronounced topological features in the optimized liquid cooling plate. More complex structural topology requires the FGTO to complete one iteration over a longer time period. However, since the FGTO method reduces regions of poor convergence, fewer iterations are needed for similar structural complexities. Consequently, the total computational times for the FGTO at $s$=3000 and $s$=5000 are decreased by 19.1% and 0.5%, respectively, compared to the conventional TO, while increased by 39.5% at $s$=8000, as shown in **Table 5**. Given that the FGTO at $s$=8000 enables better thermal performance, the additional time cost is conceivable.

The structural topology of the optimized liquid cooling plates can be manipulated by varying $s$, thus manipulating thermal and hydraulic performances. According to the fractal geometry theory [39], the recommended manipulation range for $s$ is $10^2$–$10^4$. If $s$ is too small, it fails to meet the fundamental self-similarity criterion of the fractal geometry theory. If *s* is too large, as shown in **Fig. 6**, the effect of varying *s* on the sensitivity gradually levels-off, while an extremely large *s* would amplify the convergence deterioration issue caused by extreme thermal weights, which is further discussed in section 3.3. Within the manipulable range of *s*, i.e., $10^2$–$10^4$, smaller *s* yields better manipulation effect but results in smaller solid-liquid sensitivity differences, hindering separation of solid and liquid phases and escaping from local optimal solutions. Larger *s* is conducive to obtaining more complex structures, but the manipulation effect brought by varying *s* is weakened.

### 3.2 Performance evaluation of the FGTO results

The 3D numerical calculations are performed at the inlet temperature of 308.15 K, inlet flow velocity of 0.2 m/s (*Re*=1000) and heat flux of 10 W/cm² to investigate the performances of the optimized liquid cooling plates at various *s*. The FGTO results at *s*=1000, 3000, 5000 and 8000 are selected as representatives to demonstrate the impact of *s* on the performance of the optimized liquid cooling plate. The rectangular cooling plate (RCP) with flow channel and solid wall widths of 2 mm, as shown in **Fig. S1**, is used as a benchmark for performance comparison with the FGTO results. The Nusselt number $Nu = \frac{hD_h}{k_f}$ and the performance evaluation criterion $PEC = \frac{Nu/Nu_r}{(\Delta P/\Delta P_r)^{\frac{1}{3}}}$ are used for evaluating the performances of liquid cooling plate.

**(a)**

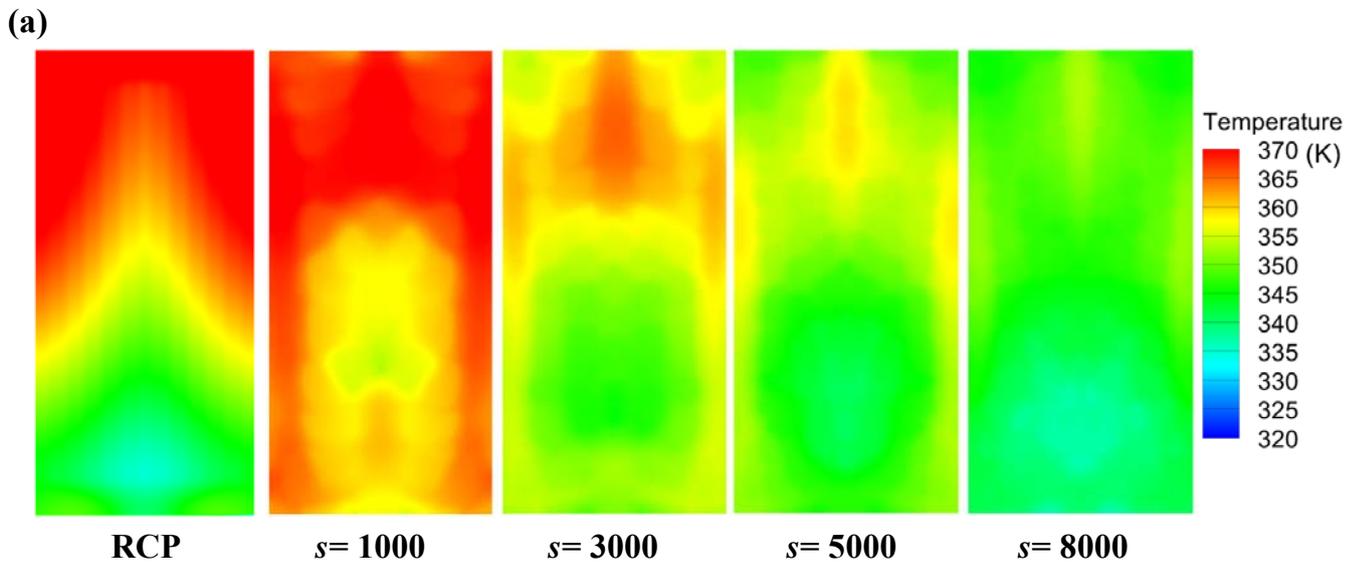

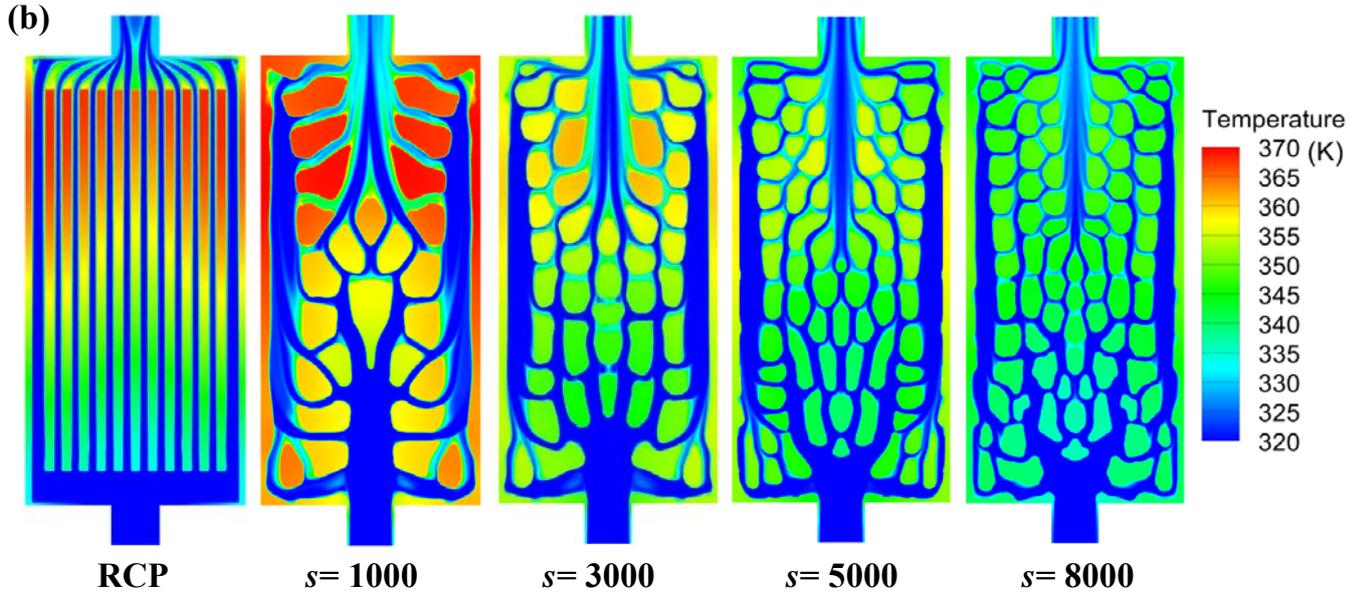

**Fig. 9.** The temperature contours of different liquid cooling plates at $T_{in}$=308.15 K, $u_{in}$=0.2 m/s and heat flux of 10 W/cm². (a) temperature contours of the bottom surface on the heated plate, (b) temperature contours of the middle plane on the direction of thickness.

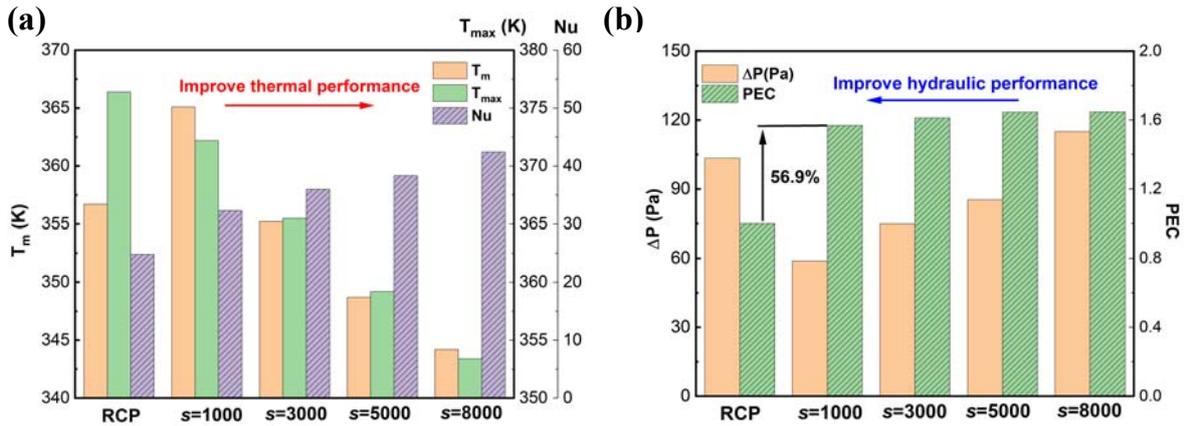

**Fig. 10.** The results of the 3D performance evaluation at $T_{in}$=308.15 K, $u_{in}$=0.2 m/s and heat flux of 10 W/cm². (a) average temperature and maximum temperature of the bottom surface and the Nusselt number, (c) pressure drop and the PEC

**Fig. 9** shows the temperature contours on the bottom and middle planes of the RCP and the FGTO liquid cooling plates at various $s$, where it can be observed that the highest temperatures occur at the ends and edges of the liquid cooling plates and the thermal performance of the liquid cooling plate is improved as $s$ increases, resulting in lower temperatures and enhanced temperature uniformity.

**Fig. 10(a)** shows the results of the thermal performance of different liquid cooling plates in terms of the average temperature $T_m$ of the bottom surface, the maximum temperature $T_{max}$, and the Nusselt number $Nu$. At $s$=3000, 5000, and 8000, the average temperatures are reduced by 1.5 K, 8 K, and 12.5 K respectively compared to the RCP, and the maximum temperatures are reduced by up to 10.9 K, 17.2 K, and 23 K, respectively, as shown in **Fig. 10(a)**. Larger $s$ leads to a more complex structural topology, enhancing flow perturbation and augmenting heat transfer area, thereby improving the heat dissipation capability of the optimized liquid cooling plates. At $s$=8000, the *Nu* of the optimized liquid cooling plates is 31.3% higher than that at $s$=3000, and 70.9% higher than that of the RCP. Thus, varying $s$ can manipulate thermal performance effectively. However, such a manipulation effectiveness gradually diminishes as $s$ increases because the impact of $s$ on intensifying solid-liquid sensitivity differences weakens with increasing $s$, as shown in **Fig. 6**. The average temperature at $s$=3000 is 9.9 K lower than that at $s$=1000, while the average temperature at $s$=8000 is only 4.5 K lower than that at $s$=5000.

**Fig. 10(b)** shows the results of hydraulic and overall performances of various liquid cooling plates in terms of the pressure drop *ΔP* and PEC. It is demonstrated that the pressure drop gradually increases while the PEC remains essentially unchanged as $s$ increases. Increasing $s$ introduces a greater flow channel tortuosity, enhancing flow perturbation while simultaneously increasing flow resistance. For the FGTO results at $s$=1000, 3000, and 5000, the pressure drops are reduced by up to 45.9%, 28.5%, and 17.5% respectively compared with those of the RCP, indicating that varying $s$ can also effectively manipulate hydraulic performance. At $s$=8000, despite of an 11.2% higher pressure drop, the thermal performance of the optimized liquid cooling plate is significantly better than that of the RCP, with the average and maximum temperatures reduced by up to 12.5 K and 23 K, respectively. At $s$=1000, although the average temperature is 8.4 K higher, flow resistance is lower and temperature distribution is more uniform compared to the RCP, with the maximum temperature reduced by 4.2 K. Consequently, its PEC, which characterizes overall performance, is up to 56.9% higher than that of the RCP.

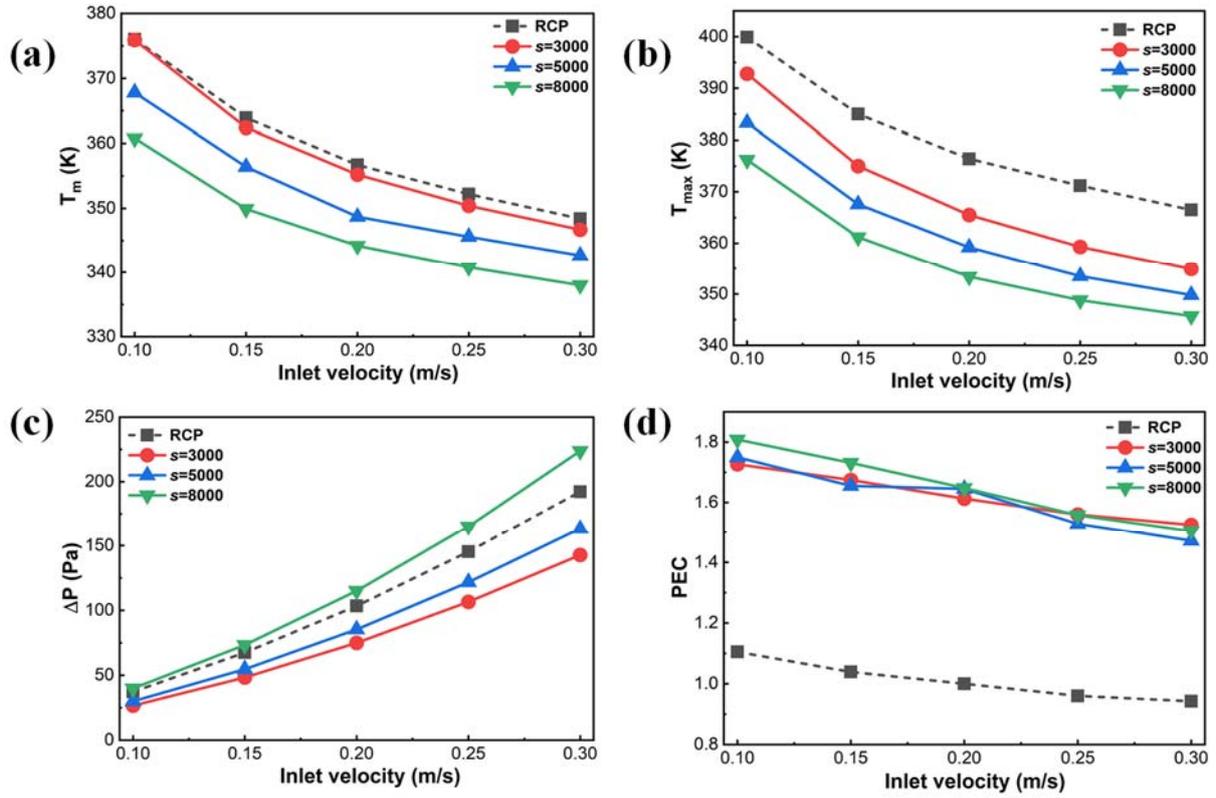

**Fig. 11.** Variation of the average temperature, maximum temperature, pressure drop and the PEC with inlet velocity. (a) average temperature, (b) maximum temperature, (c) pressure drop, (d) the PEC

**Fig. 11** shows the variations of $T_m$, $T_{max}$, $\Delta P$ and the PEC of the optimized liquid cooling plates and the RCP with the inlet velocity. It is demonstrated that the optimized liquid cooling plates by the FGTO method demonstrates outstanding thermal and hydraulic performances under varying operating conditions. For the FGTO result at $s$=5000 and inlet velocity of 0.1 m/s, the average and maximum temperatures are 8.2 K and 16.6 K lower than those of the RCP, respectively, while the pressure drop is reduced by up to 24.5%. At the inlet velocity of 0.3 m/s, the average and maximum temperatures are reduced by 5.7 K and 16.7 K respectively, while the pressure drop is reduced by 17.8%. The PEC of different liquid cooling plates decreases with the inlet velocity due to the rising pressure drop, as shown in **Fig. 11(d)**. For the optimized liquid cooling plates at $s$=5000, the PEC is 58.3% higher than that of the RCP at the inlet velocity of 0.1 m/s, while it is 56.4% higher at the inlet velocity of 0.3 m/s, indicating superior overall performance of the FGTO liquid cooling plates under different operating conditions. It is further demonstrated that varying $s$ can effectively manipulate the thermal and hydraulic performances at different inlet velocities, as shown in **Fig. 11**, suggesting strong adaptability of the FGTO method to various operating conditions. Additionally, the FGTO can effectively improve thermal and hydraulic performances not only under the uniform heat source conditions, but also under a more realistic non-uniform heat source condition, as shown in **Fig. S9**. At $Re$=1000 and local heat flux $q$=50 W/cm$^2$, the average temperatures of the heating zones of the FGTO result at $s$=8000 and $w_t$=0.7, can be reduced by 15.1 K and 14.4 K, respectively, compared to those of the RCP and the conventional TO at $w_t$=0.7.

### 3.3 Comparison of the conventional TO and FGTO at varying $w_t$

The FGTO method is an enhancement based on the conventional TO. Therefore, the FGTO can achieve a more flexible and efficient performance manipulation not only by varying $s$, as discussed in earlier sections, but also by varying $w_t$ as in the conventional TO [25]. This section compares the effect of varying $w_t$ on the FGTO and conventional TO, followed by a discussion of the synergistic effect of varying $s$ and $w_t$ on the FGTO.

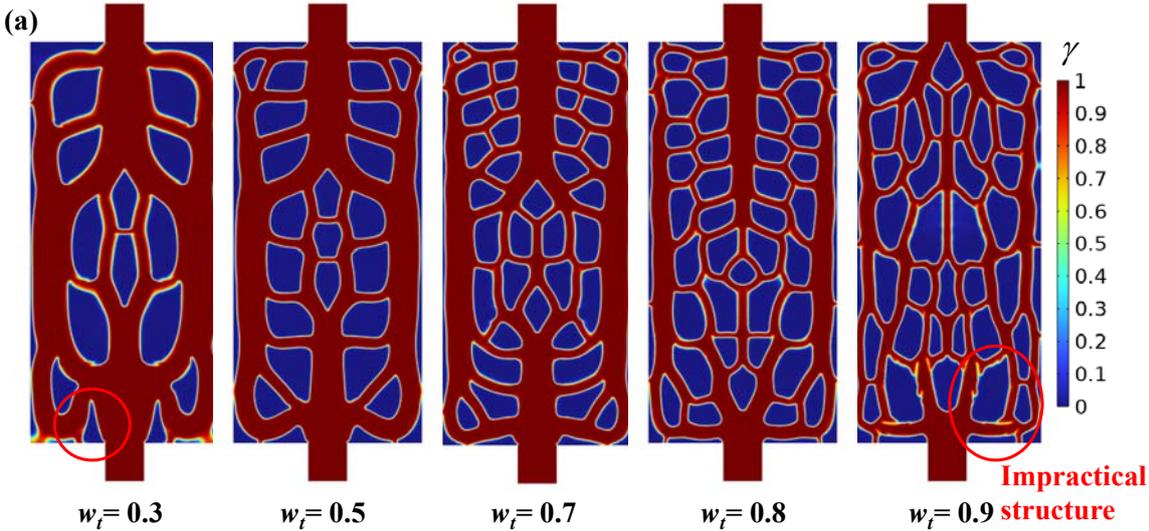

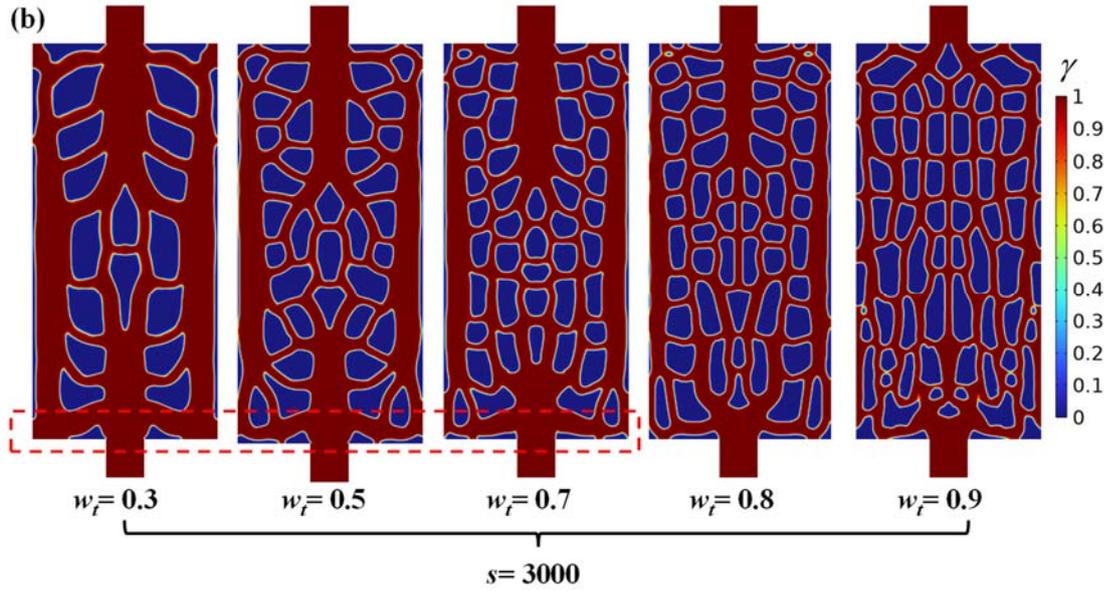

**Fig. 12.** Comparison of the effects of $w_t$ on the conventional TO and FGTO. (a) design variable field of the conventional TO results at various $w_t$, (b) design variable field of the FGTO results at various $w_t$ at $s=3000$

**Fig. 12(a)** shows the optimization results at various $w_t$ by the conventional TO. It can be observed that under moderate $w_t$, e.g., 0.5-0.8, large solid bulks break into smaller pieces as $w_t$ increases, resulting in more flow channel branches and an augmented heat transfer area. However, when $w_t$ is extremely increased to 0.9, larger solid bulks and narrower flow channels are generated in the optimization results nevertheless with impractical burr-like structures, indicating deteriorated convergence emerge, as circled in **Fig. 12(a)**. At extremely low $w_t$ =0.3, an excessively protruding flow-diverting structure forms near the inlet due to the overemphasis on hydraulic objective, as circled in **Fig. 12(a)**, which hinders rational flow distribution to the sides. These above issues stem from the inherent defect of the simple weighted-sum method of multi-objective optimization [31, 63]. At extreme $w_t$, i.e., $w_t$ approaches 1 or 0, the optimization would be easily trapped in local optimal solutions, overly emphasizing a single thermal or hydraulic objective while severely compromising the counterpart. Consequently, the optimization becomes inefficient and performance manipulation by varying $w_t$ is compromised, yielding only a slight improvement in heat dissipation capacity at the cost of extremely large flow resistance and even fails to further improve thermal performance [25]. Consequently, for the conventional TO at $w_t$ =0.7, extremely increasing $w_t$ to 0.9 can only augment the heat transfer area by 8.5%, as shown in **Table 6**. Therefore, the conventional TO struggles to adapt to scenarios with extremely high thermal demands such as the power electronics.

**Fig. 12(b)** shows the optimization results at various $w_t$ by the FGTO at $s=3000$. It can be observed from **Fig. 12** that the FGTO (**Fig. 12(b)**) achieves a larger heat transfer area and more complex structural topology at any $w_t$ than those of the conventional TO (**Fig. 12(a)**). Since the weighted-sum method is also adopted in the FGTO for bi-objective optimization, the inherent issues associated with extreme $w_t$ cannot be entirely avoided. However, compared to the conventional TO, these issues can be significantly mitigated due to better performance of the FGTO in escaping from local optimal solutions, as discussed in section 3.1. Consequently, the FGTO at $s$=8000 achieves an augmented heat transfer area improvement by up to 46.8%, 46.6% and 45.5% at $w_t$ =0.7, 0.8 and 0.9, respectively, as shown in **Table 6**.

**Table 6** Comparison of the lateral heat transfer area of the liquid cooling plates by the conventional TO and FGTO.

| $A_s$ (mm$^2$) | $w_t$=0.7 | $w_t$=0.8 | $w_t$=0.9 | Improvement by increasing $w_t$ |
|---|---|---|---|---|
| Conventional TO | 5025.9 | 5386.4 | 5453.4 | +8.5% |
| FGTO: $s$=3000 | 5681.1 | 6326.2 | 6556.7 | +15.4% |
| FGTO: $s$=8000 | 7380.5 | 7896.7 | 7935.8 | +7.5% |
| Improvement by FGTO | +46.8% | +46.6% | +45.5% | +57.9% |

*The improvement in the column direction is calculated by the FGTO at $s$=8000 and the conventional TO, while it in row direction by the results at $w_t$=0.9 and $w_t$=0.7. The improvement of 57.9% is calculated by the FGTO at $s$=8000 and $w_t$=0.9, and the conventional TO at $w_t$=0.7.

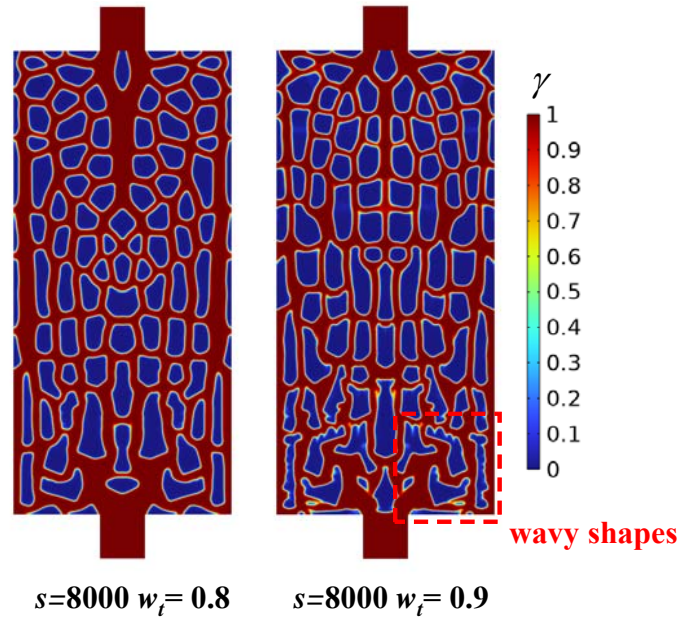

$s$=8000 $w_t$= 0.8    $s$=8000 $w_t$= 0.9

**Fig. 13.** The FGTO results at $s$=8000, $w_t$=0.8 and $s$=8000, $w_t$=0.9

As discussed earlier, increasing either $s$ or $w_t$ alone can enhance heat dissipation capabilities of the FGTO liquid cooling plates at the cost of higher flow resistance.

Simultaneously increasing both $s$ and $w_t$ would increase heat transfer area and structural complexity more significantly, leading to a further improved thermal performance. **Fig. 13** shows the FGTO results at the high $w_t$ of 0.8 and 0.9 while at the high $s$ of 8000. Compared with the FGTO results at $s$=3000, as shown in **Fig. 12(b)**, it can be observed that increasing $s$ can further enhance structural complexity even at the extremely high thermal weights. By increasing $s$ and $w_t$ simultaneously to $s$=8000 and $w_t$=0.9, the FGTO can improve heat transfer area by up to 57.9% compared to the conventional TO at $w_t$=0.7, as shown in **Table 6**. However, since both large $w_t$ and large $s$ leads to a complex structural topology that cause heavy computational costs, the regions with convergence difficulties might emerge at extreme $w_t$ as well as extreme $s$, e.g., numerous fragmented solids with arbitrary wavy shapes are formed at $s$=8000 and $w_t$=0.9, as squared in **Fig. 13**. Therefore, although the FGTO method can significantly increase the upper limit for optimizing heat transfer area, it remains constrained by the poor convergence at both high $s$ and high $w_t$.

**Fig. 14** shows the temperature contours on the bottom surface of the FGTO results at $s$=8000 and the conventional TO results at various $w_t$. It is demonstrated that the FGTO significantly enhances thermal performance of the optimized liquid cooling plates under various $w_t$, and both the FGTO and the conventional TO achieve the best thermal performance at around $w_t$=0.8. Compared with the optimal thermal performance by the conventional TO at $w_t$=0.8, the FGTO at $s$=8000 and $w_t$=0.8 further reduce the average and maximum temperatures by up to 15.6 K and 16.9 K, respectively, as shown in **Fig. 14**. Consequently, the FGTO significantly elevates the upper limit of thermal performance optimization.

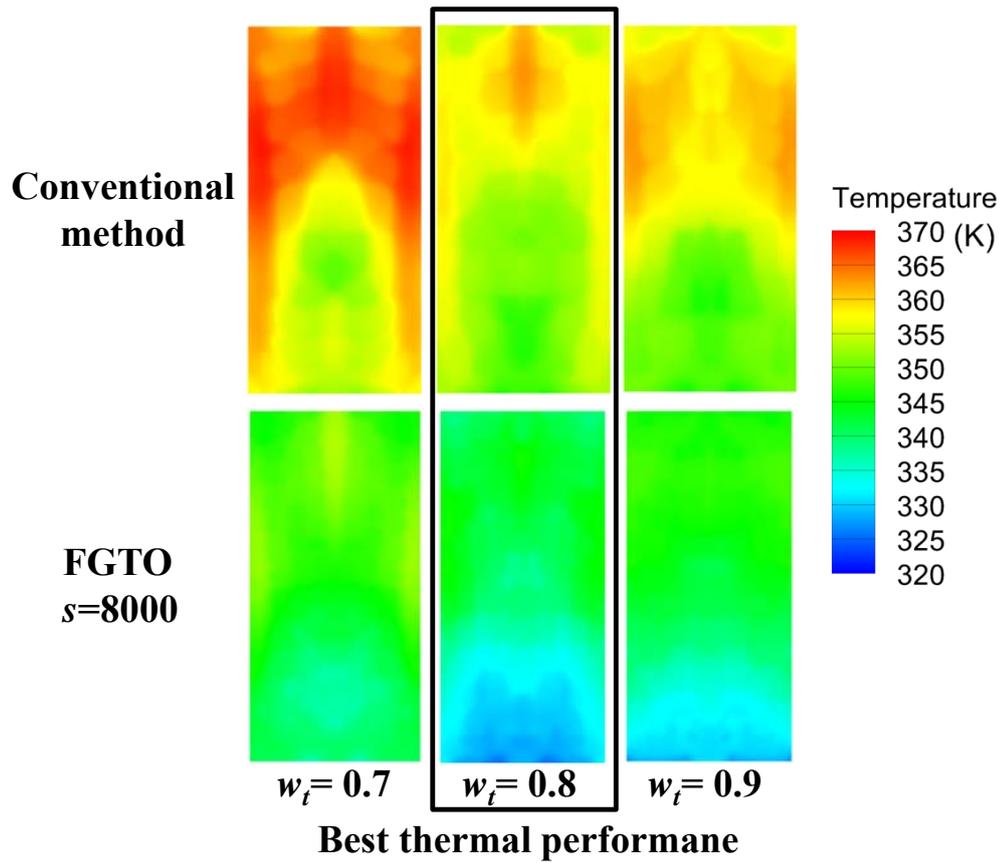

**Fig. 14.** The temperature contours on the bottom surface of the FGTO result at $s$=8000 and the conventional TO result at various $w_t$, with inlet velocity of 0.2 m/s, inlet temperature of 308.15 K and heat flux of 10 W/cm$^2$

## 3.4 Performance manipulation by the conventional TO and FGTO

As discussed in section 3.3, simply varying $w_t$ by the conventional TO struggles to efficiently manipulate thermal and hydraulic performances of the optimized liquid cooling plates due to the inherent limitations of the weighted-sum method. Since the FGTO enables performance manipulation by varying $s$ at a moderately constant $w_t$, as discussed in section 3.1, it circumvents the issues arising from the extreme thermal weights, leading to a more efficient performance manipulation.

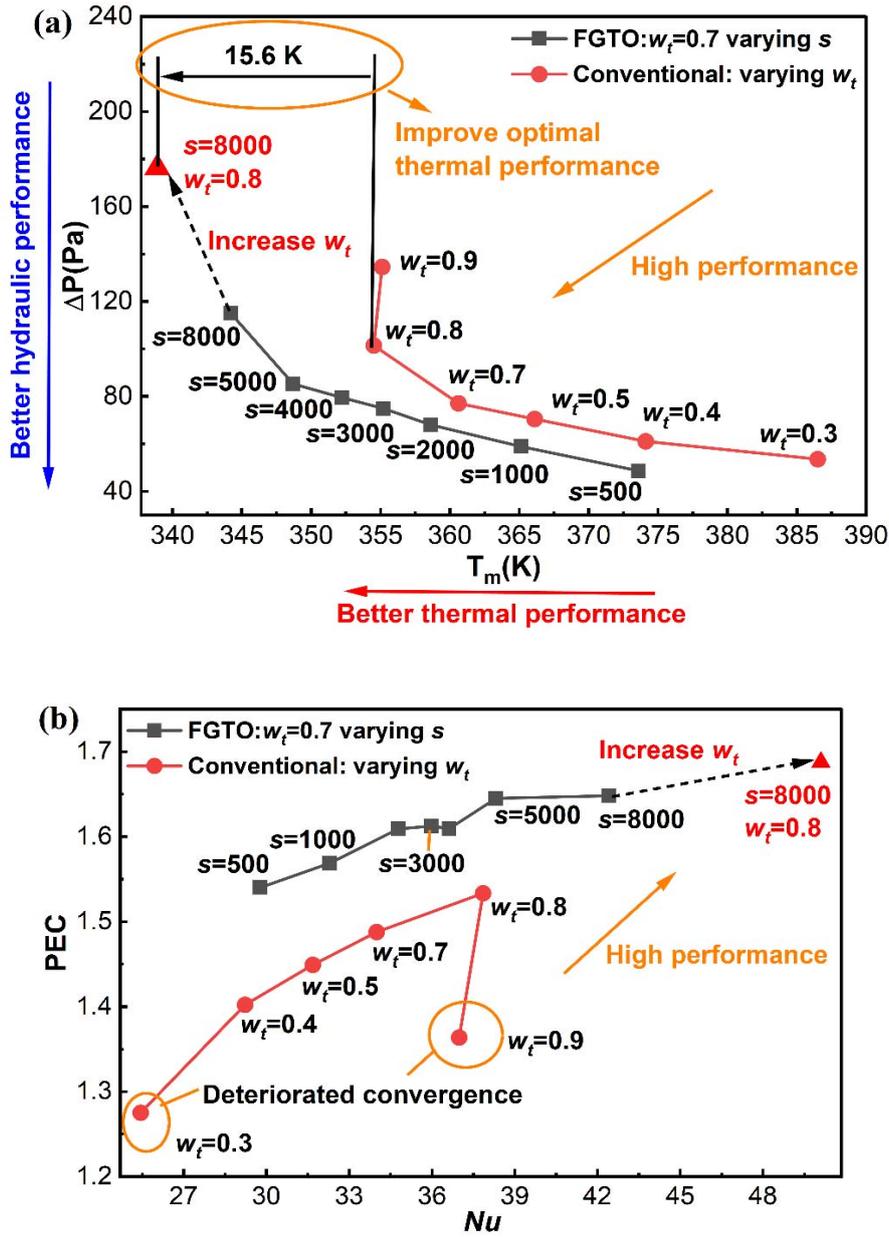

**Fig. 15.** Performance comparison of the optimized liquid cooling plates by the FGTO at $s$=500, 1000, 2000, 3000, 4000, 5000, 8000 at $w_t$ =0.7 and by the conventional TO at varying $w_t$=0.3, 0.4, 0.5, 0.7, 0.8, 0.9, under the operating condition of $T_{in}$=308.15 K, $u_{in}$=0.2 m/s and heat flux of 10 W/cm². (a) average temperature and pressure drop, (b) the Nusselt number and the PEC

**Fig. 15** shows the comparison of performance manipulation through varying $s$ by the FGTO at $w_t$=0.7, and through varying $w_t$ by the conventional TO. It is demonstrated that the FGTO enables more efficient performance manipulation compared with the conventional TO. By varying $s$ at a moderately constant $w_t$=0.7, the FGTO can achieve superior thermal performance at a comparable hydraulic performance, and vice versa, as shown in **Fig. 15(a)**. For example, the average temperature of the FGTO result at $s$=5000 is 5.8 K lower than that of the conventional TO result at $w_t$=0.8, while the pressure drop is reduced by 15.9%. The pressure drop of the FGTO result at $s$=500 is about 9.1% lower than that of the conventional TO result at $w_t$=0.3, while the average temperature can be reduced by up to 12.9 K.

For the scenarios with extremely high thermal demands, where the conventional TO fails to further improve thermal performance at $w_t$=0.8, the FGTO can further reduce the average temperature by up to 15.6 K by increasing $s$ and $w_t$ simultaneously, as shown by the black dotted arrow in **Fig. 15(a)**. Meanwhile, despite of a pressure drop rise of 74.8 Pa, the overall performance indicated by the PEC of the FGTO at $s$=8000 and $w_t$=0.8 is 10.1% higher than that of the conventional TO result at $w_t$=0.8, while the *Nu* is also improved by 32.2%, as shown in **Fig. 15(b)**. Furthermore, the PEC of the conventional TO is significantly compromised at extremely high or low $w_t$ due to the deteriorated convergence, whereas the FGTO enables efficient performance manipulation while maintaining a high PEC at various $s$, as shown in **Fig. 15(b)**.

Overall, increasing the input parameter *s* can effectively enhance thermal performance of the optimized liquid cooling plate at the cost of a larger pressure drop. As shown in **Fig. 15(b)**, the overall performance indicated by the PEC is improved with rising *s*, suggesting that the benefits from improved heat dissipation capabilities outweigh the drawbacks from the insignificant rise in pressure drop as *s* is increased. Therefore, it is recommended to raise parameter *s* close to its upper limit in practical applications to achieve a better overall performance. As discussed in Section 3.1, the fractal geometry theory requires that *s* be effectively manipulated within the range of $10^2$ to $10^4$. Given that as *s* approaches its upper limit of $10^4$, the effect of varying *s* on objective function sensitivity and consequently on the FGTO results gradually levels off, as shown in **Fig. 6** and **Fig. 5(b)**, it allows considerable flexibility in selecting an optimal *s* value, e.g., *s*=8000 is recommended for practical applications. Although simultaneously increasing *s* and $w_t$ can achieve a further improvement in thermal performance, the convergence deterioration issue would emerge at extreme *s* and $w_t$ values. Compromising the performance improvement and computational robustness, a moderate $w_t$ should be selected while increasing the *s*. For instance, *s*=8000 and $w_t$=0.8 is considered an optimal parameter combination in this study, which achieves the best thermal performance and PEC in the optimized liquid cooling plates as shown in **Fig. 15**.

### 3.5 Performance comparison of the FGTO and other previous works

The FGTO results are compared with those by conventional TO methods using different objective functions. One is the result of the bi-objective of heat generation formulated as Eq.

(10) and the pumping power [24]. The other is the result by the triple objectives of average solid temperature, maximum temperature, and pumping power [64]. The geometries and boundary conditions used in the 2D optimization and 3D numerical calculations are exactly the same as those in the respective original studies, as shown in **Fig. S3** and **Fig. S4**, and the corresponding thermo-physical properties are listed in **Table S4** and **Table S5**.

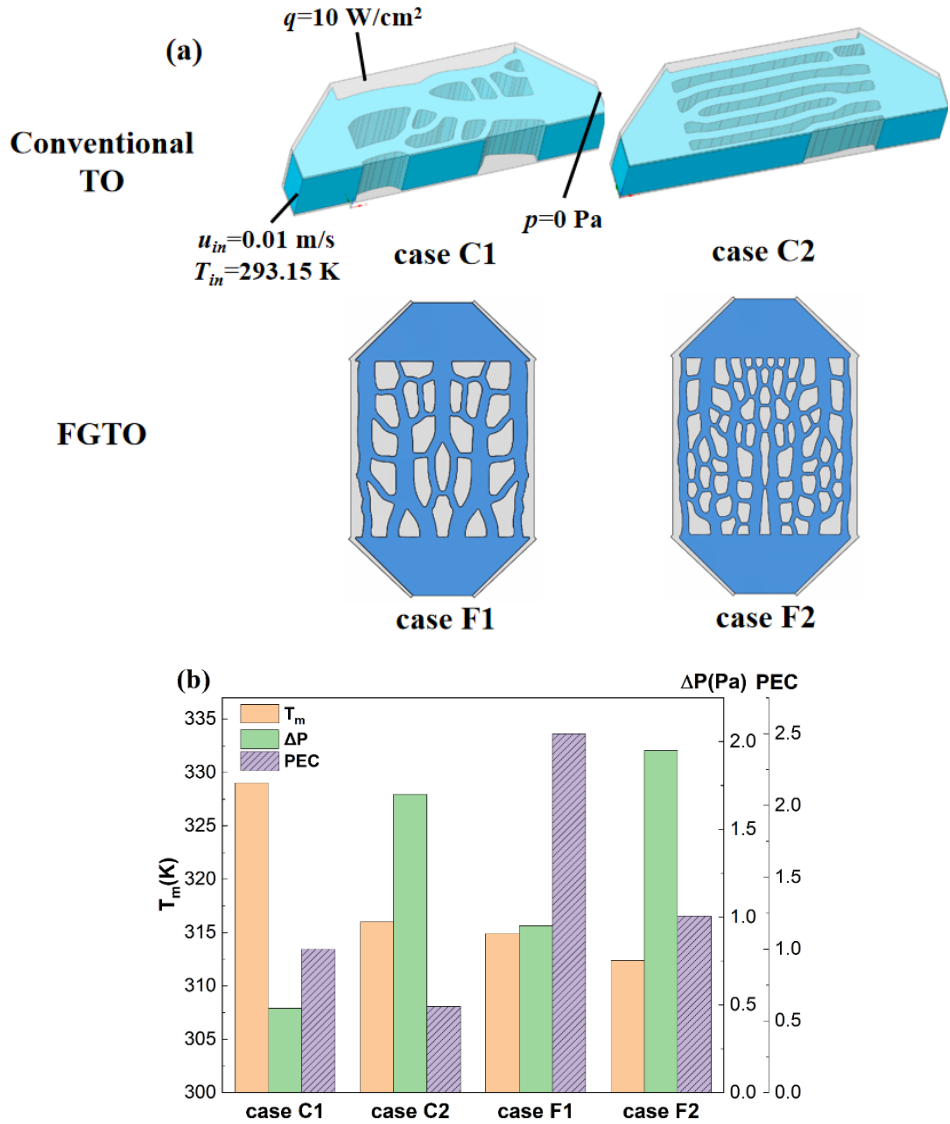

**Fig. 16.** Performance comparison of the liquid cooling plates by the FGTO method and conventional TO method in [24]. (a) the FGTO results and the conventional TO results, (b) average wall temperature, pressure drop and the PEC of the FGTO results and conventional TO results

**Fig. 16(a)** shows the optimization results in [24] and the present study, with the labels "C" and "F" referring to the results obtained by the conventional TO and FGTO, respectively. **Fig. 16(b)** shows the average temperature $T_m$, pressure drop $\Delta P$, and the PEC of the conventional TO and FGTO liquid cooling plates. The structural topology and thermal performance of the FGTO results can be manipulated by varying $s$, as shown in **Fig. S5** and **Fig S6**. By optimizing at $s$=1000 (case F1), the average temperature can be reduced by 14.1 K, while the PEC is improved by 60%, compared with that of case C1. By optimizing at $s$=8000 (case F2), the average temperature is reduced by 3.6 K compared with that of case C2, while the PEC is increased by 51.2%, as shown in **Fig. 16(b)**. Additionally, the FGTO results exhibit superior thermal and overall performances under varying operating conditions, as shown in **Fig. S7**.

**Fig. 17(a)** shows the TO results in [64] obtained by the conventional method (case C3 and case C4) and the results by the FGTO method at $s$=5000 (case F3). The 2D optimization is performed at the inlet velocity of 0.1 m/s, and the 3D numerical calculation is performed at the inlet flow rate ranges from 300 mL/min to 600 mL/min, with both heat fluxes of 1 W/cm². **Fig. 17(b)** shows the variation of the maximum temperature of the top plate with the inlet flow rate, and the temperature over the entire plate is shown in **Fig. S8**. At the flow rate of 300 mL/min, the maximum temperature $T_{max}$ of the FGTO result is reduced by 2.8 K, compared to that in [64]; while at the flow rate of 600 mL/min, $T_{max}$ is reduced by 3.0 K. **Fig. 17(c)** and **Fig. 17(d)** show the variations of the pressure drop and PEC with inlet flow rate, respectively. The FGTO result also demonstrates a superior hydraulic performance, with the pressure drop of 26% lower than that of the conventional TO result at the flow rate of 600 mL/min. Consequently, the FGTO result can achieve better overall performance under various operating conditions, with the PEC increased by 38.4% at the flow rate of 300 mL/min, and increased by 15% at the flow rate of 600 mL/min, compared to those by the conventional TO.

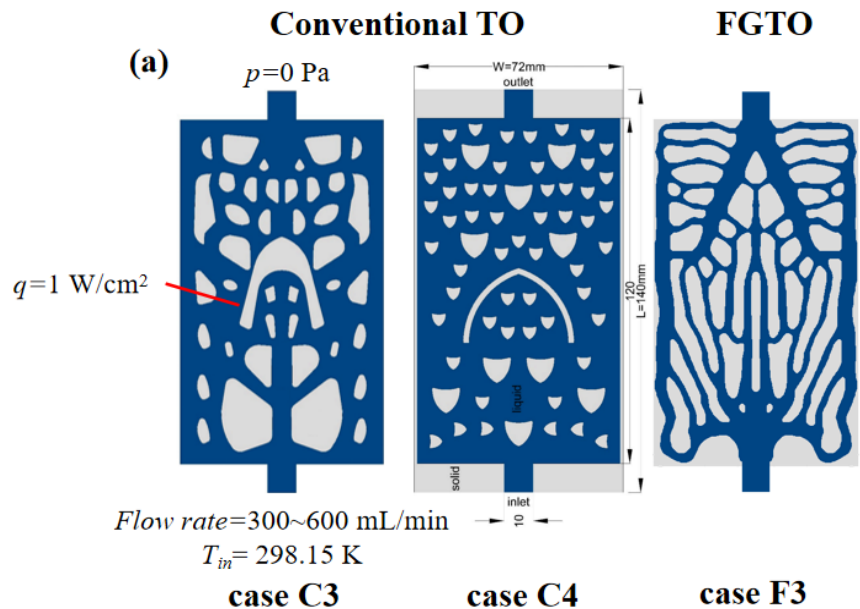

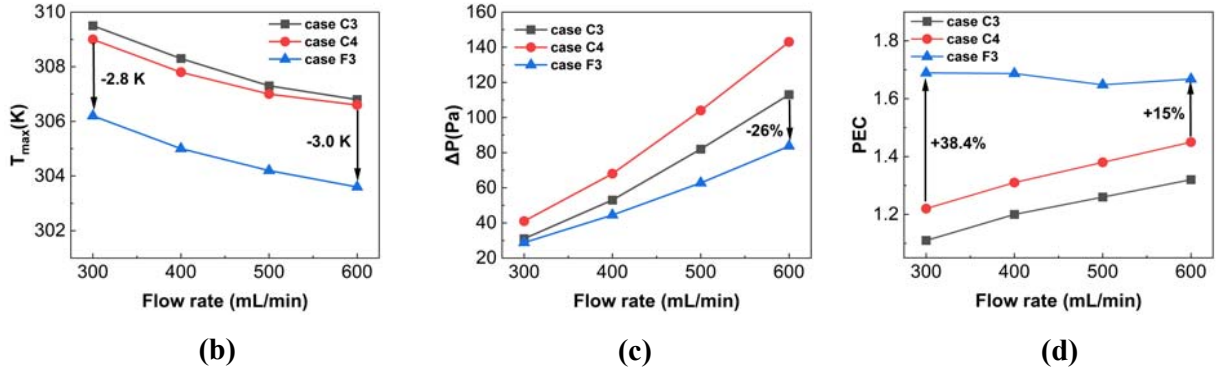

|  (b)  |  (c)  |  (d)  |

**Fig. 17.** Comparison of the conventional method in [64] and the FGTO method. (a) the TO results obtained by the conventional and FGTO methods, (b) variation of the maximum temperature of the bottom plate with flow rate, (c) variation of the pressure drop with flow rate, (d) variation of the PEC with flow rate

### 3.6 Feasibility and adaptability of the FGTO method

Although the FGTO method can yield the optimized liquid cooling plates with enhanced structural complexity, its outstanding convergence ensures high computational efficiency. For the most FGTO in this study, convergence can be achieved within 200 iterations, as shown in **Table 5**. Compared with the conventional TO with comparable structural topology, the FGTO enables shorter computational time for convergence. Furthermore, the FGTO method also exhibits broad adaptability to the thermo-physical properties of various solids and coolants. As shown in **Fig. 18**, the FGTO method can be effectively applied to both lower solid-liquid thermal conductivity ratios $K$ and higher coolant viscosities $\mu$, which demonstrates feasibility of applying the FGTO method to liquid cooling plates with different coolants such as oil, phase change slurry and so on. The FGTO improves the density-based TO by explicitly depicting the topological features of porous media using fractal geometry theory, which is not only applicable to liquid cooling plates design, but might also be extended to other scenarios where the density-based TO are applicable, such as fins design for air cooling or energy storage.

Although the fragmented solid structures generated by the FGTO at large $s$ might pose challenges for manufacturing, it remains feasible for practical application. The minimum length scale in the optimized liquid cooling plates can be effectively controlled by appropriately increasing the filter radius. Additionally, the advancements in additive manufacturing techniques enhance the manufacturing precision of complex structural topologies, e.g., current direct metal laser sintering technique can reliably process flow channels with hydraulic diameters of 400–600 μm [65], which essentially meets the manufacturing requirements for the FGTO liquid cooling plates.

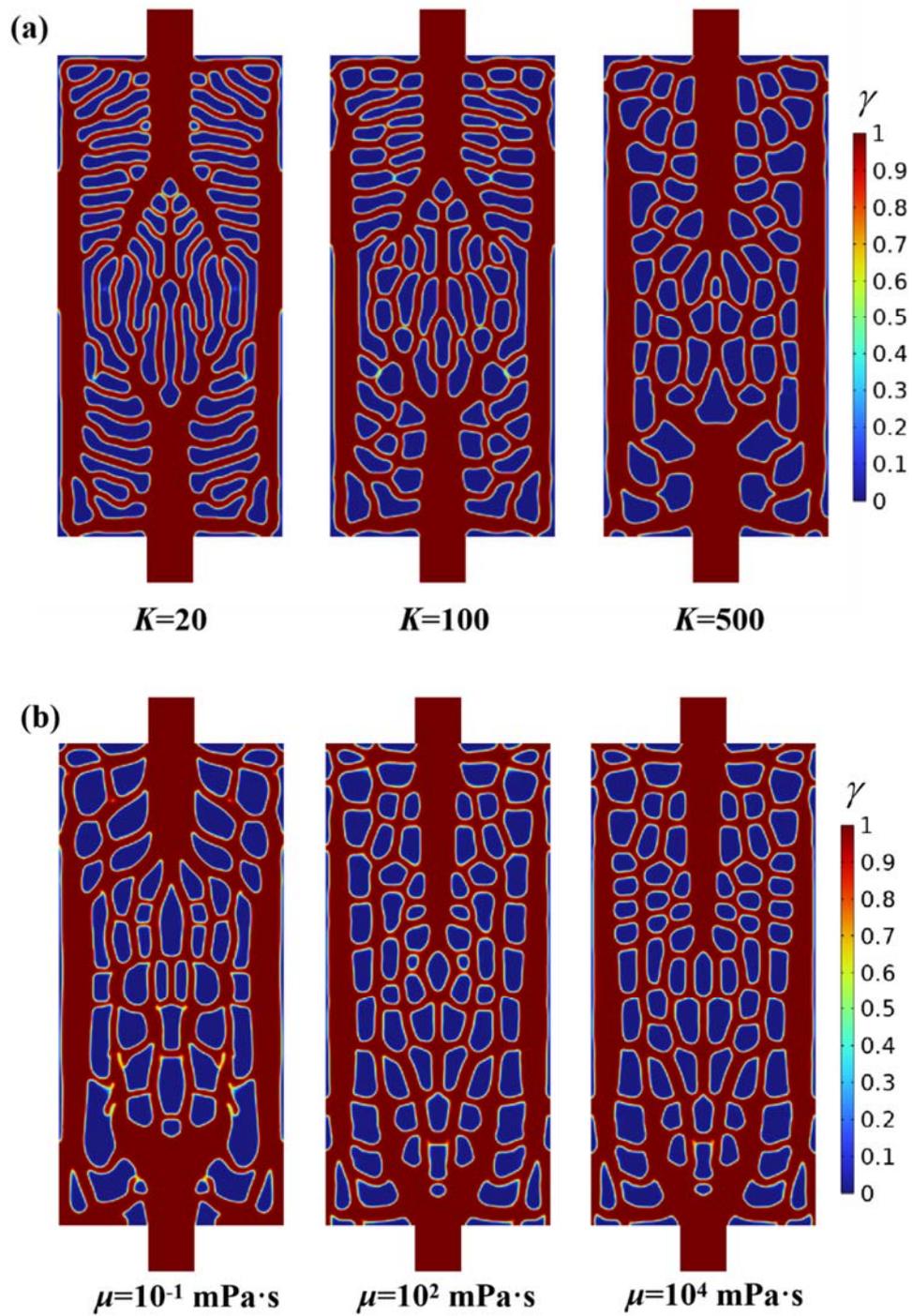

**Fig. 18.** The FGTO results for different solid to liquid thermal conductivity ratios $K$ and coolant viscosities $\mu$ at $s$=5000 and $w_t$=0.7, at inlet velocity of 0.01 m/s, inlet temperature of 308.15 K and heat flux of 1 W/cm$^2$. (a) design variable field at various $K$, (b) design variable field at various $\mu$.

## 4. Conclusion

This study proposes a fractal geometry topology optimization (FGTO) method for liquid cooling plate design. The FGTO explicitly depicts the heat transfer area through the fractal geometry theory to achieve a direct optimization of convective heat transfer in the thermal objective function, and introduces an input parameter $s$ that is related to fractal dimension to manipulate the optimization results. The performance of the FGTO results is evaluated through the 3D numerical calculations at the inlet velocity of 0.2 m/s, inlet temperature of 308.15 K and a heat flux of 10 W/cm$^2$. The main conclusions are as follows:

(1) Compared to the RCP, maximum temperature of the FGTO results can be reduced by up to 23 K and the pressure drop can be reduced by 45.9%. Compared to the conventional TO, the FGTO achieves a more complex structural topology in the optimized liquid cooling plates and the heat transfer area is improved by up to around 46%. The FGTO raises upper limit of the thermal performance, with the average temperature reduced by up to 15.6 K compared to the optimized results by the conventional TO, while the PEC improved by 10.1% simultaneously.

(2) The FGTO achieves a more efficient performance manipulation by varying $s$ at moderately constant $w_t$, thus circumventing the issue of performance degradation at extreme $w_t$ by the conventional TO. Compared to varying $w_t$ by the conventional TO, varying $s$ by the FGTO achieves a more significant thermal performance improvement while sacrificing less hydraulic performance. Increasing $s$ to an extreme value while fixing $w_t$ at a relatively moderate value can achieve optimal thermal and overall performances while ensuring the computational robustness and convergence, e.g., $s=8000$ and $w_t=0.8$ is recommended as an optimal parameter combination.

(3) The mechanism the FGTO achieving a more complex structural topology lies in that it explicitly depicts heat transfer area of the porous media by fractal geometry theory and directly optimizes convective heat transfer rather than the heat conduction by the conventional TO. It is mathematically manifested that the FGTO enhances sensitivity differentiation of the objective functions between solid and liquid phases, which facilitates solid-liquid separation from the optimized porous media and is conducive to escaping from the local optimal solutions.

## CRediT authorship contribution statement

Zixu Han: Writing – original draft, Methodology, Investigation, Formal analysis. Kairan Yang: Validation. Peng Zhang: Writing – review & editing, Validation, Supervision, Funding acquisition, Conceptualization.

**Declaration of competing interest**

The authors declare that they have no known competing financial interests or personal relationships that could have appeared to influence the work reported in this paper.

**Data availability**

Data will be made available on reasonable request

**Acknowledgements**

This research is supported by the Natural Science Foundation of Shanghai Municipality under the Contract No. 25ZR1401209. This research is also partially supported by the National Natural Science Foundation of China under the Contract No. 52576220. A few characterizations are conducted at the AEMD of Shanghai Jiao Tong University.